\documentclass[12pt,preprint]{aastex}
\usepackage{graphicx}
\usepackage{amssymb}
\usepackage{amsmath}
\newcommand{\be}{\begin{equation}}
\newcommand{\ee}{\end{equation}}

\shorttitle{Convergence of halo properties} 
\shortauthors{Trenti et al.}

\begin{document}


\title{How well do cosmological simulations reproduce individual-halo properties?}


\author{Michele Trenti\footnote{trenti@colorado.edu}, Britton D. Smith, Eric J. Hallman, Samuel W. Skillman, J. Michael Shull} 
\affil{University of Colorado, CASA, Dept.  of Astrophysical \& 
Planetary Sciences, 389-UCB, Boulder, CO 80309, USA} 


\begin{abstract}
  Cosmological simulations of galaxy formation often rely on
  prescriptions for star formation and feedback that depend on halo
  properties such as halo mass, central over-density, and virial
  temperature. In this paper we address the convergence of individual
  halo properties, based on their number of particles $N$, focusing in
  particular on the mass of halos near the resolution limit of a
  simulation. While it has been established that the halo mass
  function is sampled on average down to $N \sim 20-30$ particles, we
  show that \emph{individual} halo properties exhibit significant
  scatter, and some systematic biases, as one approaches the
  resolution limit. We carry out a series of cosmological simulations
  using the Gadget2 and Enzo codes with $N_p=64^3$ to $N_p=1024^3$
  total particles, keeping the same large-scale structure in the
  simulation box.  We consider boxes of small ($l_{box} =
  8~\mathrm{Mpc~h^{-1}}$), medium ($l_{box} = 64
  ~\mathrm{Mpc~h^{-1}}$) and large ($l_{box} = 512
  ~\mathrm{Mpc~h^{-1}}$) size to probe different halo masses and
  formation redshifts. We cross-identify dark matter halos in boxes at
  different resolutions and measure the scatter in their properties.
  The uncertainty in the mass of single halos depends on the number of
  particles (scaling approximately as $N^{-1/3}$), but the rarer the
  density peak, the more robust its identification. The virial radius
  of halos is very stable and can be measured without bias for halos
  with $N \gtrsim 30$. In contrast, the average density within a sphere
  containing 25\% of the total halo mass is severely underestimated
  (by more than a factor 2) and the halo spin is moderately
  overestimated for $N\lesssim 100$. If sub-grid physics is
  implemented upon a cosmological simulation, we recommend that rare
  halos ($\sim 3\sigma$ peaks) be resolved with $N \gtrsim100$
  particles and common halos ($\sim 1\sigma$ peaks) with $N \gtrsim
  400$ particles to avoid excessive numerical noise and possible
  systematic biases in the results.
\end{abstract}


\section{Introduction}\label{sec:intro}

Cosmological simulations are a fundamental tool for investigating the
formation and evolution of dark matter halos and of the resulting
galaxies \citep{bert98}. With advancements in both memory and
computing power capabilities, simulations can now be performed with up
to several billion particles. Efforts have been focused both on
investigating the formation of single halos with ultra-high resolution
\citep[e.g., ][]{vialactea,aquarius,stadel09} and on simulating
structure formation in large boxes, on the order of Gpc$^3$, that
represent the large-scale structure in the Universe. \citep[e.g.,
][]{mill,marenostrum}. However, the dynamic range of scales to be
resolved is so large, despite the continuous hardware improvements,
that numerical simulations need to be pushed to their limits for many
interesting problems. For example, to follow the formation of the
first stars in the Universe, a sub-solar mass resolution is needed
\citep{oshea07,turk09}. This limits the size of the simulation box to
below $\sim$ 1~Mpc$^3$, and thus large-scale structure effects are
missed. At other range of the spectrum, simulations of the formation
of bright (and thus rare) high-redshift quasars observed in all-sky
surveys such as the Sloan Digital Sky Survey \citep{fan06} require
computational volumes $\geq 10^9$ Mpc$^3$ and resolution sufficient
to identify the black hole seeds from metal-free stars, for a total
dynamic range $ \geq 10^{12}$ \citep{ts07a,tss08}. Under such conditions,
information from barely resolved halos needs to be used. In addition,
sub-grid physics recipes are often employed, either in the form of
semi-analytical (post-processing) modeling
\citep{kauffmann98,somerville99,delucia07} or as star formation and
feedback recipes implemented during the run
\citep{springel03,oppen08}.

Therefore, convergence and validation of the numerical methods used
are of fundamental importance to establish the reliability of the
conclusions drawn from numerical experiments. Past investigations have
addressed two fundamental issues regarding dark matter (DM) halos: the
convergence of the inner slope of the density profile and the accuracy
of the halo mass function. The inner-slope problem, arising from the
absence of observational evidence of the density cusps predicted in
simulations \citep{NFW}, has been the focus of several studies using
extremely high resolution \citep{ghigna00,power03,fukushige04}, but the
issue appears to be settled, with the inner regions expected to follow
a Einasto density profile \citep{navarro08,stadel09}.
Regarding the DM halo mass function, it is now well established that a
cosmological simulation reproduces with fidelity the mass function
down to halos with $N \gtrsim 20-30$ particles (\citealt{reed03}; see
also \citealt{heitmann06,warren06,lukic07}), although there can be
box-size effects \citep{bagla05,reed07,reed09}. These results are also
in excellent agreement with analytical predictions
\citep{st99,jenkins01}.

As we will demonstrate, the mass function is only an average property
of the halo mass distribution. A much deeper question pertains to the
reliability of properties of single halos, derived from a simulation
with given resolution, { with a goal similar to the study of
subhalos properties in the Aquarius run \citep{aquarius,springel08b}.}
{ Individual halo} properties affect the variance of the
results derived from the simulations (\citealt{warren06}). For
example, if semi-analytical formulae and/or star formation recipes are
implemented in a run, the uncertainty on individual halos
propagates to the derivation of quantities such as the star formation
rate, the fundamental plane thickness \citep{fp87} or the tightness in
the relation between the central black hole mass and bulge velocity
dispersion\citep{ferrarese00}. 

The goal of this paper is to quantify the numerical scatter and
identify possible biases in the mass of individual halos as a function
of the number of particles in the halo. Past investigations have
characterized the behavior of halo finders for idealized systems with
small N, for example by generating discrete realizations of a
\citet{NFW} profile or by downsampling the resolution of a simulation
snapshot \citep{warren06,lukic09}. We extend these studies by studying
the convergence of halo properties in a fully cosmological context,
where the convergence properties of the N-body integration are also
investigated. We carry out a suite of simulations, where higher
resolution boxes are constrained to have the same phases as
low-resolution versions. We cross-identify halos between the different
runs and measure convergence of their properties, which turns out to
be different with respect to the more idealized numerical experiments
of \citet{warren06} and \citet{lukic09}. Our study should help the
community of numerical cosmologists to quantify the limit at which
they should trust their simulations, depending on the desired accuracy
goal.

This paper is organized as follows. In Section.~\ref{sec:num} we
introduce the setup of our suites of simulations, whose results are
presented in Section.~\ref{sec:result}. We conclude in
Section.~\ref{sec:conc} by discussing our recommendations for the
minimum resolution required for implementing extra physics on dark
matter halos.

\section{Numerical setup}\label{sec:num}


We generate initial conditions using the Grafic1 package
\citep{bertschinger01}, with a custom modification that allows us to
apply the \citet{hoffman91} method over the full simulation box and to
use the \citet{eisenstein99} power-spectrum fitting formula. With our
customization, we are able to start from a low-resolution version of
the initial conditions and then refine it to higher resolution while
keeping the same large-scale structure. We use a WMAP-5 cosmology
\citep{komatsu08} with $\Omega_{\Lambda} = 0.72$, $\Omega_m = 0.28$,
$\Omega_b = 0.0462$, $\sigma_8=0.817$, $n_s = 0.96$ and $h=0.7$. We
consider three different box sizes: a small box of edge
$l_{box}=8~\mathrm{Mpc}~{h^{-1}}$, a medium box (edge $l_{box}=64~
\mathrm{Mpc}~{h^{-1}}$ ) and a large box ($l_{box}=512~
\mathrm{Mpc}~{h^{-1}}$). The simulations have a range of total particles
from $N_p=64^3$ to $N_p=1024^3$ (the number of particles in a single
halo is instead indicated as $N$). The small-box simulations start at
redshift $z=199$, while the medium and large-box simulations begin at
$z=100$. The boxes have periodic boundary conditions. Details on the
specific simulations, including their mass resolution, are shown in
Table~\ref{tab:sim}.


As our code of choice we use the particle-mesh tree code Gadget2 in
its ``lean'' version \citep{springel05,mill} to carry out dark-matter
only simulations. The softening parameter is set to $\epsilon =
l_{box}/(42.5 N_p)$, allowing us to achieve a good spatial resolution
of virialized halos with a small number of particles. For comparison,
we also carry out a subset of the runs (the medium-box series) using
the hydrodynamic code Enzo \citep{bryan95}.
Enzo\footnote{http://lca.ucsd.edu/projects/enzo} uses the
block-structured adaptive mesh refinement (AMR) scheme of
\citet{berger1989} to achieve high spatial and temporal resolution,
and it combines an N-body adaptive particle-mesh solver for dark
matter dynamics with a Piecewise Parabolic Method (PPM) hydro solver
that has been optimized for cosmological applications
\citep{woodward1984,bryan95}. Because our primary goal is to test the
DM halo dynamics of Enzo, we do not include gas in the Enzo
simulations. The number of top-grid cells in our Enzo runs is equal to
the total number of particles. 

The main difference between the Gadget2 and Enzo under these
conditions is their spatial resolution. The force resolution in Enzo
is twice the grid size, that is $2 l_{box}/N_p^{1/3}$. In Gadget2, the
force becomes unsoftened at a distance of about $3 \epsilon$, or at
$\sim l_{box}/ (14 N_p^{1/3})$ in our simulations. The Gadget2 runs
thus have about $30$ times better spatial resolution than their Enzo
counterparts, if a uniform grid and no AMR is used in Enzo. To
investigate the effect of force resolution on the properties of halos,
we have carried out a subset of the Enzo runs ($N_p=64^3$ to
$N_p=256^3$) allowing up to six levels of AMR. This improves the force
resolution by a factor up to $2^6$, thereby reaching a maximum force
accuracy comparable to that attained by Gadget2. Enzo still has a
lower force resolution in regions with overdensities below the
critical threshold for AMR (see \citealt{oshea05}). { While many
  applications of Enzo rely on aggressive use of AMR, for example in
  the context of the formation of Population III stars
  \citep{oshea07,turk09}, a growing number of investigations consider runs with uniform resolution
  \citep{unigrid3,unigrid2,unigrid1,unigrid0}. Especially for studies
  of the Ly$\alpha$ forest, it has been shown by \citet{unigrid3} that
  disabling AMR provides an order-of-magnitude speedup, while only
  introducing $\sim 5\%$ errors in the Ly$\alpha$ flux power spectrum.}

We save snapshots of the simulations at regular redshift intervals
($z=6,4,1,0$) and we identify DM halos with a friends-of-friends (FoF)
algorithm \citep{davis85} with linking length 0.2. We also analyze a
subset of runs with the Amiga halo finder \citep{knebe08} that
includes a boundness check for the halo particles. In addition, the
Amiga halo finder provides detailed information on each halo,
including their density profile and spin parameter. 

We consider halos with at least 32 particles. To cross-identify the
same halo in two simulations at different resolution, we match
individual particle identification numbers (IDs) that are
representative of the initial particle positions. { Our method is
  similar to that discussed in \citet{aquarius}}. The ID of every
particle in the lower resolution realization (with $N=N_{low}$) is
used to calculate its corresponding $n_c = N_{high}/N_{low}$
``child''\footnote{If the $N_p$ is increased by $2^3$, then each
  low-resolution particle has $2^3$ counterparts, or ``children'' at
  high-resolution. The average position and velocity of the
  ``children'' corresponds approximately, but not exactly, to the
  position and velocity of their low-resolution ``parent''
  \citep{bertschinger01}.} particle IDs in the high-resolution
simulation (with $N=N_{high}$). From the list of particle IDs in each
halo we can thus ascertain whether that halo has one or more
counterparts in the higher-resolution run. Similarly, given a halo in
the high-resolution simulation, we can determine the presence of any
low-resolution counterparts. Note that the relation between halos in
snapshots at two different resolutions is not necessarily one-to-one
or one-to-zero. In fact, multiple halos can be the counterparts of a
single larger halo, especially in the process of merging. For any
given low-resolution halo, we identify its high-resolution counterpart
by considering the high-resolution halo that has the largest number of
individual ``children'' particle matches. In passing we note that if
matching of halos is based instead on their positions, there is
no guarantee of either positive or unique identification. This
affects especially common low-mass objects in the proximity of larger
halos, because a change in resolution can lead to different tidal
forces and changes in the rate of mergers.

\section{Results}\label{sec:result}

\begin{figure}
\includegraphics[scale=0.4]{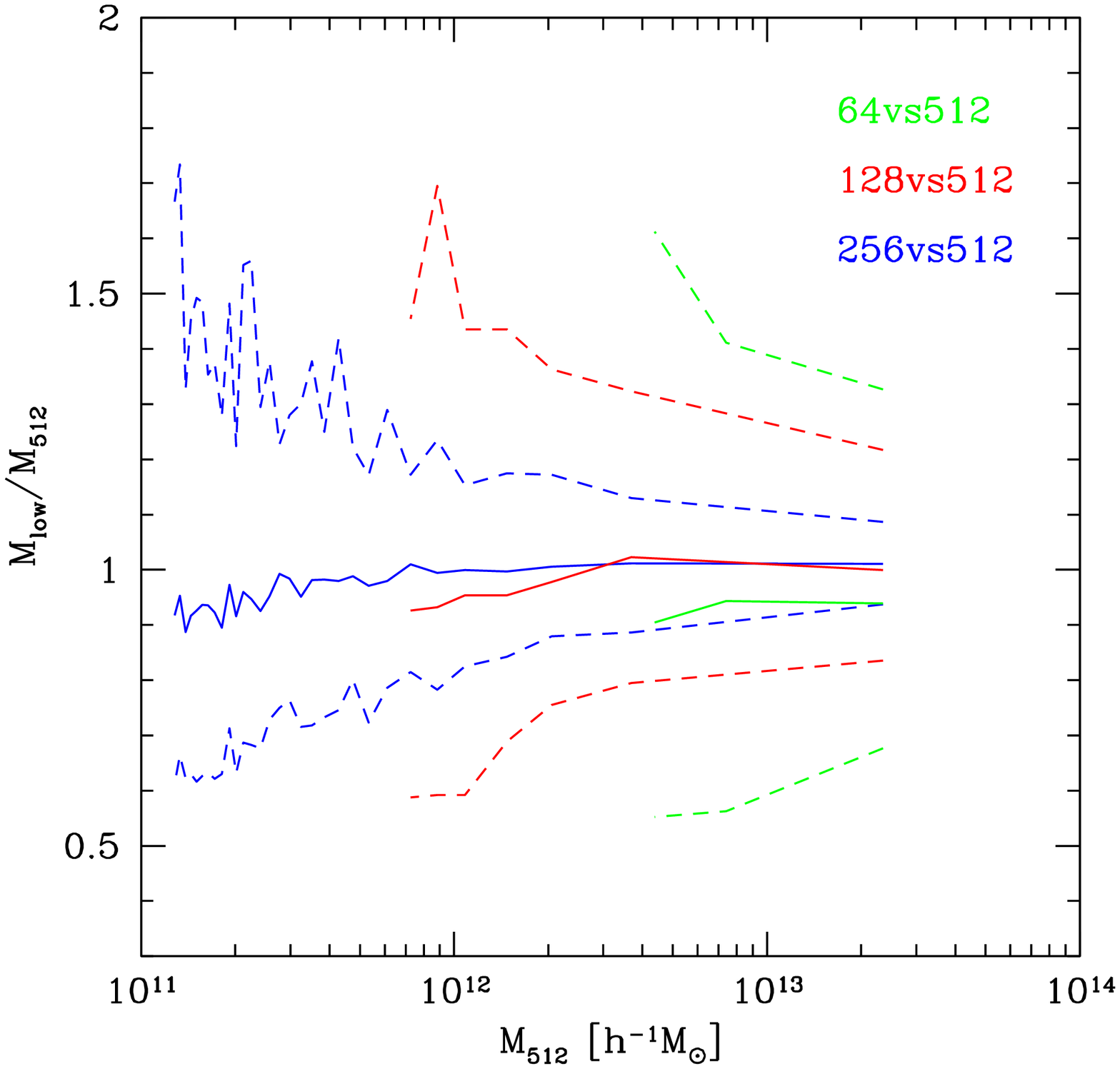}
\includegraphics[scale=0.4]{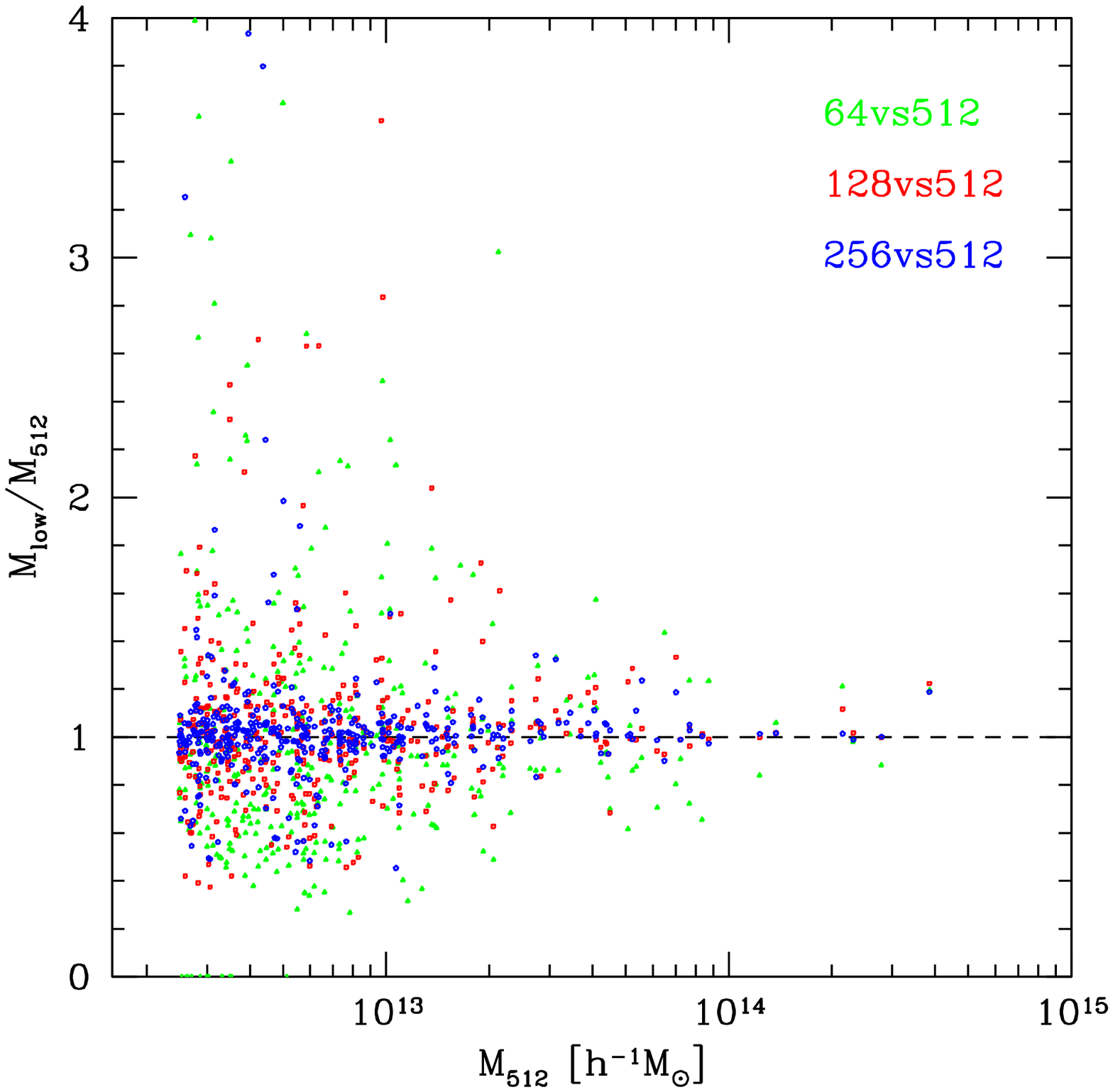}
\caption{{ Comparison of individual halo masses at $z=0$ for four
    realizations of the same initial conditions (medium box:
    $l_{box}=64~\mathrm{Mpc~h^{-1}}$) at different resolution, from
    $N_p=64^3$ to $N_p=512^3$. The two panels show the ratio
    $M_{low}/M_{512}$ of halo mass measured at low resolution against
    the value found in the highest resolution run as a function of the
    halo mass at high resolution. Halos have been identified in the
    high-resolution realization ($N_p=512^3$) and matched to
    counterparts at lower resolution. Left panel shows the median of
    $M_{low}/M_{512}$ as a solid line. Dashed
    red lines delimit the symmetric $1\sigma$ region around median. 
Color coding is as follows. Green: $N_{low}=64^3$; red: $N_{low}=128^3$; blue: 
$N_{low}=256^3$. Right panel: zoom into the individual $M_{low}/M_{512}$ 
values for the most massive halos (identified in all realizations).}}
\label{fig:convergence} \end{figure}

In Figure~\ref{fig:convergence} we show the scatter of individual halo
masses in our medium-box simulation at $z=0$ when the resolution {
  progressively increases from $N_p=64^3$ to $N_p=512^3$. We plot the
  ratio of low-to-high resolution mass of halos $M_{low}/M_{512}$ as a
  function of the halo mass in the highest-resolution run $M_{512}$.
  As the resolution is increased, it is clear that the halo mass is
  measured with progressively higher accuracy. In
  Figure~\ref{fig:mass_scatter} we show the scatter as a function of
  the number of particles in the low resolution run ($N_{low}$).} The
  median mass of
  halos is correct down to about 100 particles, while it tends to be
  underestimated for the smallest halos. However, the halo mass
  function remains consistent as shown in Figure~\ref{fig:mf} {
    (see also \citealt{reed07})}. We
  quantify the dimensionless scatter, $\xi(N_{low})$, around the
  median for halos with $N_{low}$ particles by considering one half of
  the symmetric $1\sigma$ interval that encloses from $16\%$ to $84\%$
  of the $M_{low}/M_{high}$ points for halos with $N_{low}-\Delta N
  \leq N \leq N_{low}+\Delta N$. We chose $\Delta N \gtrsim 10$
  adaptively to ensure that the distribution is well sampled. The
  scatter around the median grows steadily as the number of particle
  decreases. For halos with less than $10^3$ particles, their mass has
  $\gtrsim 20 \%$ uncertainty at 68\% confidence level. The
  convergence of the mean halo mass down to a small number of
  particles is consistent with previous resolution studies
  \citep{reed03} that demonstrated that the halo mass function of a
  simulation is correctly sampled down to such low particle number.

Although the mean halo properties we find are consistent with those
reported in earlier investigations, the individual scatter of halo
masses measured from cosmological simulations differs from the
estimates based on idealized experiments carried out by
\citet{warren06} and \citet{lukic09}. First, both \citet{warren06} and
\citet{lukic09} report that FoF halo finders tend to overestimate the
mass of discrete realizations of a halo at low $N$. In actual
simulations, the opposite behavior is observed. The mass of small $N$
halos is in fact underestimated at low resolution. By comparing the
$N_p=256^3$ realization of the medium box against the $N_p=512^3$
realization, we find that the average of $M_{low}/M_{high}$ for halos
with $32 \leq N \leq 64$ is $0.84$ and the median is $0.77$. From
Table 1 and Equation~3 of \citet{warren06} we would have instead
expected $\langle M_{low}/M_{high} \rangle \gtrsim 1.1$. We obtain a
different result because the mass of the halo in a simulation depends
not only on the convergence properties of the halo-finder algorithm,
explored by \citet{warren06} and \citet{lukic09}, but also on those of
the code that resolves the non-linear gravitational dynamics leading
to the formation of the halos. A lower number of particles, and thus a
reduced force and spatial resolution, suppresses high frequency modes
of the effective power spectrum of the simulation, producing an
underestimate of the mass of halos near the resolution limit of the
run (see also \citet{oshea05} for a similar finding in the context of
the comparison between Enzo and Gadget2). A second difference between
our findings and those published earlier is in the amplitude of the
scatter. Both \citet{warren06} and \citet{lukic09} observe in their
halo experiments a scatter in the measure of halo masses that is a
factor 2 lower than what is realized in a cosmological simulation. For
example, \citet{warren06} measure an $18\%$ relative 1-$\sigma$ error
for their synthetic $N=50$ halos. In our simulations, the 1-$\sigma$
relative error is $\gtrsim 55\%$ for halos with $N \sim 50$ as shown
in the bottom left panel of Figure~\ref{fig:mass_scatter}. Again, this
is not surprising, because our results are affected primarily by the
different resolution in the cosmological simulations, rather than by
the convergence properties of the halo finder.

Interestingly, the considerable scatter in the individual halo mass is
left essentially unchanged if the DM halo catalogs are pruned
of unbound particles when $N \gtrsim 100$. This is shown in the right
panels of Figure~\ref{fig:mass_scatter}, which are the equivalent of
the left panels but obtained using halo catalogs from the Amiga halo
finder that includes a boundness check for membership of particles to
a halo \citep{knebe08}. Removing unbound particles helps only at
the lowest end of the resolution for halos with $N \lesssim 100$; for
example $\xi_{Amiga}(N=40) \sim 0.4$ while $\xi_{FoF}(N=40) \sim
0.55$. As expected, the overall number of halos identified in a
snapshot above a given halo mass is slightly lower ($\sim 5\%$) when
unbound particles are removed. For example, in the medium box with
$N_p=512^3$ particles, there are 98,011 halos with $N \geq 32$
particles identified by the friends-of-friends halo finder and 93,364 by
the Amiga halo finder at $z=0$.

The scatter in halo masses remains largely unchanged when we consider
different box sizes and redshifts, as shown in
Figure~\ref{fig:mass_scatter5}. There is a moderate tendency for rare
halos to be better resolved at a given number of particles compared to
their more common counterparts, especially when $N \lesssim 10^2$ (see
Figure~\ref{fig:sigma_nu}). This is highlighted by quantifying the
rarity of halos using the extended Press-Schechter variable
$\nu=\delta^2_c/\sigma^2(M) $. { For example, the very common (low
  $\nu$) halos in Figure~\ref{fig:sigma_nu} have considerable more
  scatter at $N \sim 100$ than rarer halos (high $\nu$) with a similar
  number of particles}. In addition, common halos
  with $\sim 10^2$ particles may be in reality part of a larger halo
  when the numerical resolution is increased (see the points at
  $M_{\mathrm{low}}/M_{\mathrm{high}} \sim 0$ in the bottom left panel
  of Figure~\ref{fig:mass_scatter}). This effect does not happen for
  rarer halos (see the upper left panel of
  Figure~\ref{fig:mass_scatter}). This is not surprising, because if a
  halo originates from a rare peak, then it is more likely to be the
  dominant gravitational source in its surroundings and the dynamics
  of its own particles is primarily governed by self-gravity. In
  contrast, more common halos are likely to be surrounded by at least
  comparably massive neighbors, and they might be more affected by
  tidal-field errors.

The scatter of individual halo masses is reduced as the number of
particles in a halo increases, but it remains considerable, even when
a halo has $N\gtrsim 10^3$ particles (see
Figure~\ref{fig:sigma_npart}).  The scaling of convergence with the
number of particles can be understood with a simple analytical
model. For the purpose of computing the total mass of a halo, the
particles more likely to be affected by errors in their dynamics are
those initially located at the periphery of the halo over-density. In
the linear regime at $z \gg 1$, when the density field is
quasi-homogeneous, a spherical region that contains $N$ particles has
$N_{boundary}$ edge particles, where:
\begin{equation}
N_{boundary} \approx 4 \pi \left(\frac{3 N}{4 \pi}\right)^{2/3}.
\end{equation}
Assuming that a fraction of the edge particles is affected by
numerical resolution, then the dispersion of the mass of an individual
halo scales as $\xi(N) \propto N_{boundary}/N \sim N^{-1/3}$. If we
assume that, on average, about half of the edge particles are
susceptible to change of membership when the resolution is increased,
then we expect a $\sim 20 \%$ uncertainty in the mass of a halo with
$1000 \leq N \leq 2000$ particles, in reasonable agreement with the
scatter we measure ($\sim15\%$ in Figure~\ref{fig:sigma_nu}). This
means that to reduce the typical uncertainty on a halo mass below
$10\%$, $N \sim 5000$ particles are required. Because this scaling
depends only on the surface-to-volume ratio, the $N^{-1/3}$ trend is
predicted even if the spherical assumption for the collapse is relaxed
and more realistic models for the formation of halos are considered,
such as the ellipsoidal collapse model \citep{sheth01}.
Figure~\ref{fig:sigma_npart} shows that the empirical measurements for
$\xi(N)$ do indeed show that a good fit of the overall distribution is
given by:
\begin{equation}\label{eq:xi}
\xi(N) \sim 0.15 \left (\frac{N}{1000} \right)^{-1/3}.
\end{equation}

From Figures~\ref{fig:convergence} and \ref{fig:mass_scatter} it can
be seen that some halos with a large number of particles ($N\gtrsim
10^4$) can occasionally have a large variation in their mass when the
same box is resimulated at higher resolution. These are halos in the
process of merging, as shown in Figure~\ref{fig:halo_merger} for a
halo with $8192$ particles in the $l_{box}=64 \mathrm{Mpc}~{h^{-1}}$
box ($N_p=256^3$) at $z=0$. While the FoF halo finder flags the halo
as a single entity at low resolution (there is a bridge of particles
connecting the two main components), at high resolution the merging is
slightly delayed, so that the two sub-components are still separate
halos. Of course, the opposite condition may also be realized, with
two individual halos identified at low resolution and a single halo at
high resolution. Such ambiguity in defining a halo cannot be avoided
unless an additional diagnostic is used in addition to halo-finding
algorithms (such as halo profilers or indicators for an irregular
morphology). Nevertheless, the fraction of halos in the process of
undergoing a merger in any single snapshot is small, and the measure
of $\xi(N)$ is not affected by outliers in the distribution of
$M_{low}/M_{high}$.

{ The dimensionless scatter $\xi$ we measure appears larger by
  about a factor two compared to the scatter quantified for subhalos
  in the \emph{Aquarius} simulation (Figure~16 in \citealt{aquarius}; see
  also the supplementary information in \citealt{springel08b}). The
  better convergence of subhalo properties found by \citet{aquarius}
  is not surprising. In fact, subhalos are the remnants of initially
  more massive halos that have been stripped of their less bound
  particles, both during the merging with the main halo and by tidal
  forces, once the subhalo is orbiting inside the parent halo. Loosely
  bound particles are more likely to be added or removed from a halo as a result of a change in resolution. In addition, once a subhalo is part of a larger halo, its
  orbit becomes defined by the parent halo potential, and further
  merging with other subhalos is highly unlikely. Scatter in the
  individual properties of subhalos is thus not affected by the
  ambiguity in defining a halo undergoing a major merger, contrary to
  what might happen for the halos studied here
  (Figure~\ref{fig:halo_merger}).}

The convergence of individual halo properties appears worse in Enzo
runs when $N \lesssim 400$. Figure~\ref{fig:mf} shows that the halo
mass function of such runs deviates from both the analytical
\citet{st99} prediction and the higher-resolution realization when $N
\lesssim 400$. Increasing the force resolution by switching on AMR
helps with respect to Enzo Unigrid runs, but there is still a
significant number of halos missing at $N \lesssim 400$. This trend
identified in the halo mass function is also clearly visible at the
level of individual halo masses (see
Figure~\ref{fig:mass_scatter_enzo}). The median of the distribution of
$M_{low}/M_{high}$ approaches $1$ only at $N \sim 1000$, in sharp
contrast to the better convergence properties found in the Gadget2
runs. This result is not surprising and has been already previously
noted (\citealt{oshea05,hallman07}). The gravity solver in the Enzo
code does in fact suppress small-scale power at very high redshift,
before AMR refinement is triggered \citep{oshea05}. The convergence of
the halo mass can be improved by increasing the dimension of the
top-level grid, as shown by \citet{oshea05}. However, this might not
always be possible for the largest runs, when the limiting factor of
computational resources is often the availability of memory rather
than processor speed.

Figure~\ref{fig:rho25} show the convergence of other individual halo
properties, obtained by profiling the halos with the Amiga halo
finder. The core density ($\rho_{25}$), defined as the average density
within the radius containing $25\%$ of the total halo mass, is
reported in Figure~\ref{fig:rho25} for one snapshot at $z=0$ from the
medium box. { As expected, the errors on $\rho_{25}$ are larger
  than those on the total halo mass, and} there is clearly a systematic
bias: low-resolution halos with small $N$ have a significantly lower
$\rho_{25}$ compared to their realization at higher resolution. The
median of $\rho_{25~low}/\rho_{25~high}$ is as low as $\sim 0.3$ for
$N\sim 32$. This bias has a major impact if the rate of star formation
is scaled from the core density, for example by using the
\citet{schmidt59} law. The systematic underestimation of core density
in low-resolution runs is likely due to two-body relaxation at the
center of poorly resolved halos. The scatter of the virial radius,
$r_{vir}$, is shown in Figure~\ref{fig:rvir}. This quantity appears to
be well defined, and has no bias and a small scatter down to $N= 32$.
{ The stability of $r_{vir}$ is not surprising as $r_{vir} \propto
  M^{1/3}$. Hence, an error on the mass does not severely
  affect the associated radius.} Figure~\ref{fig:spin} shows the spin
of individual halos, measured by the dimensionless parameter $\lambda
= J|E|^{1/2}/GM^{5/2}$, where $J$ is the angular momentum of the halo,
$E$ its total energy (gravitational and kinetic) and $G$ the
gravitational constant. At small $N$, $\lambda_{low}$ appears
systematically higher than $\lambda_{high}$ and has a large scatter.
This is not surprising: small $N$ systems always have some residual
angular momentum, even if the particle positions and velocities were
to be drawn from the distribution function of a non-rotating system.
The bias in $\lambda$ for small $N$ may be important in semi-analytic
modeling of galaxy formation, in case the angular momentum of a halo
is used to determine the presence of a disk.

\section{Conclusions and Discussion}\label{sec:conc}

\begin{figure} 
\includegraphics[scale=0.4]{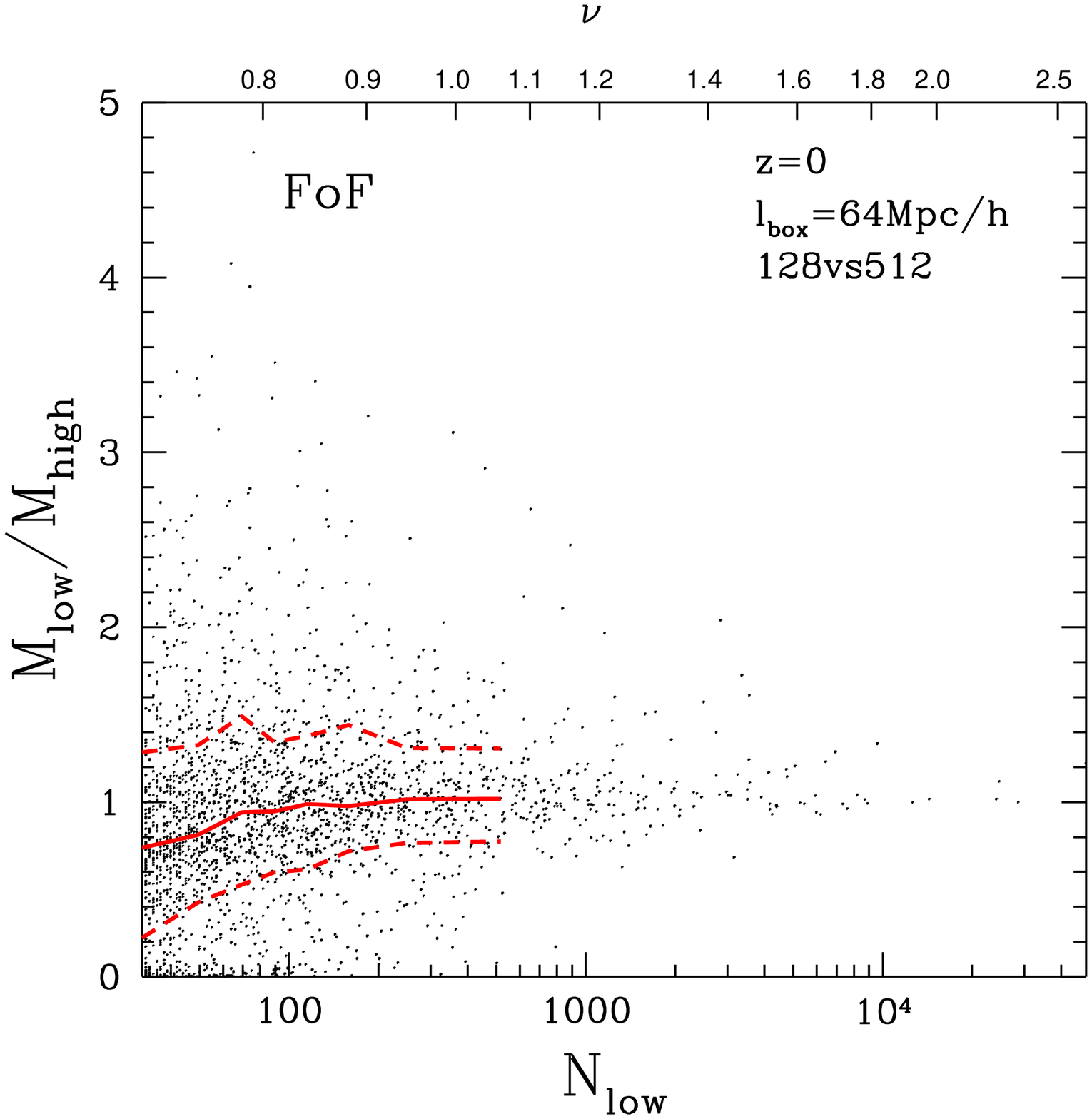}
\includegraphics[scale=0.4]{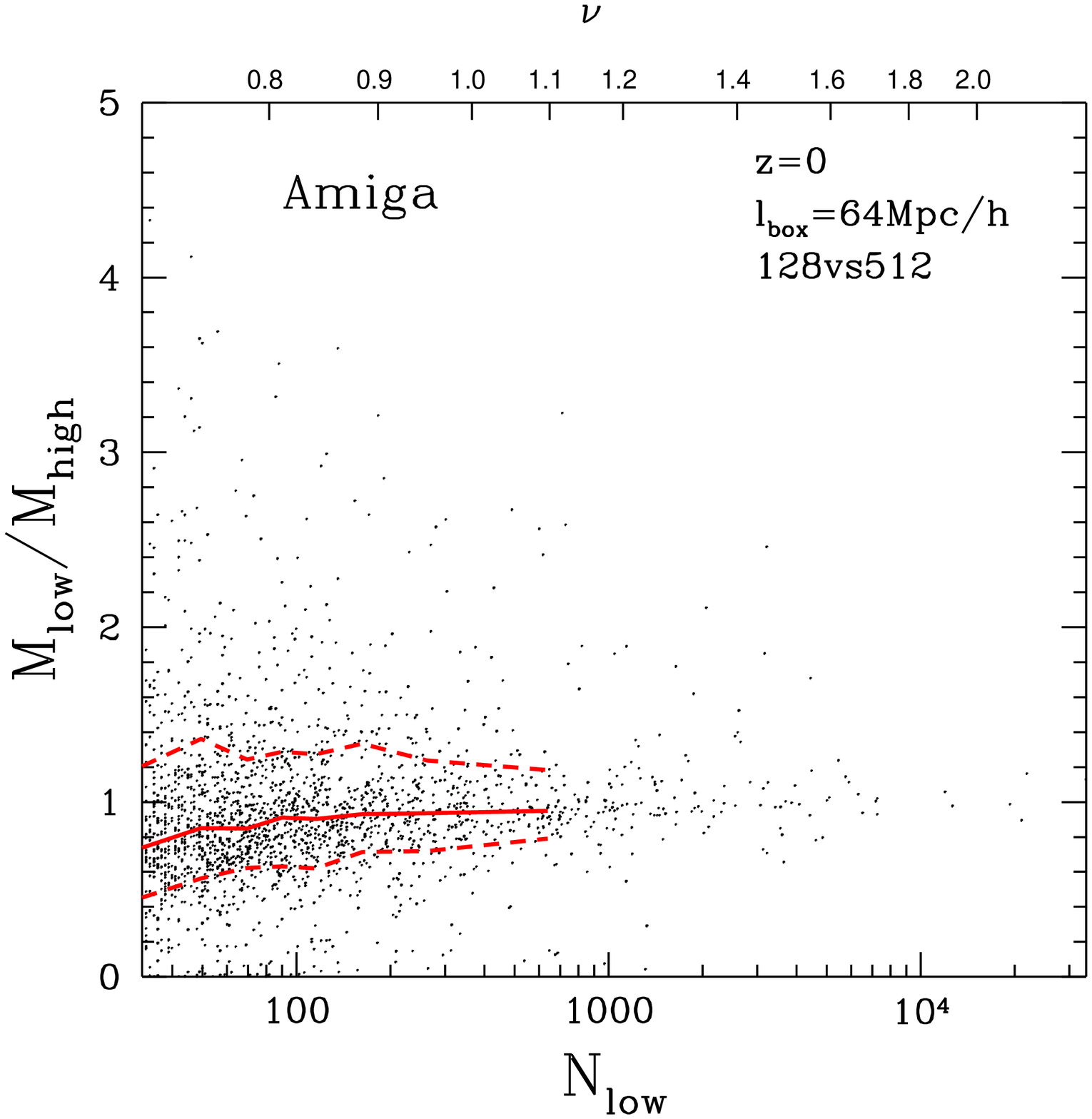}
\includegraphics[scale=0.4]{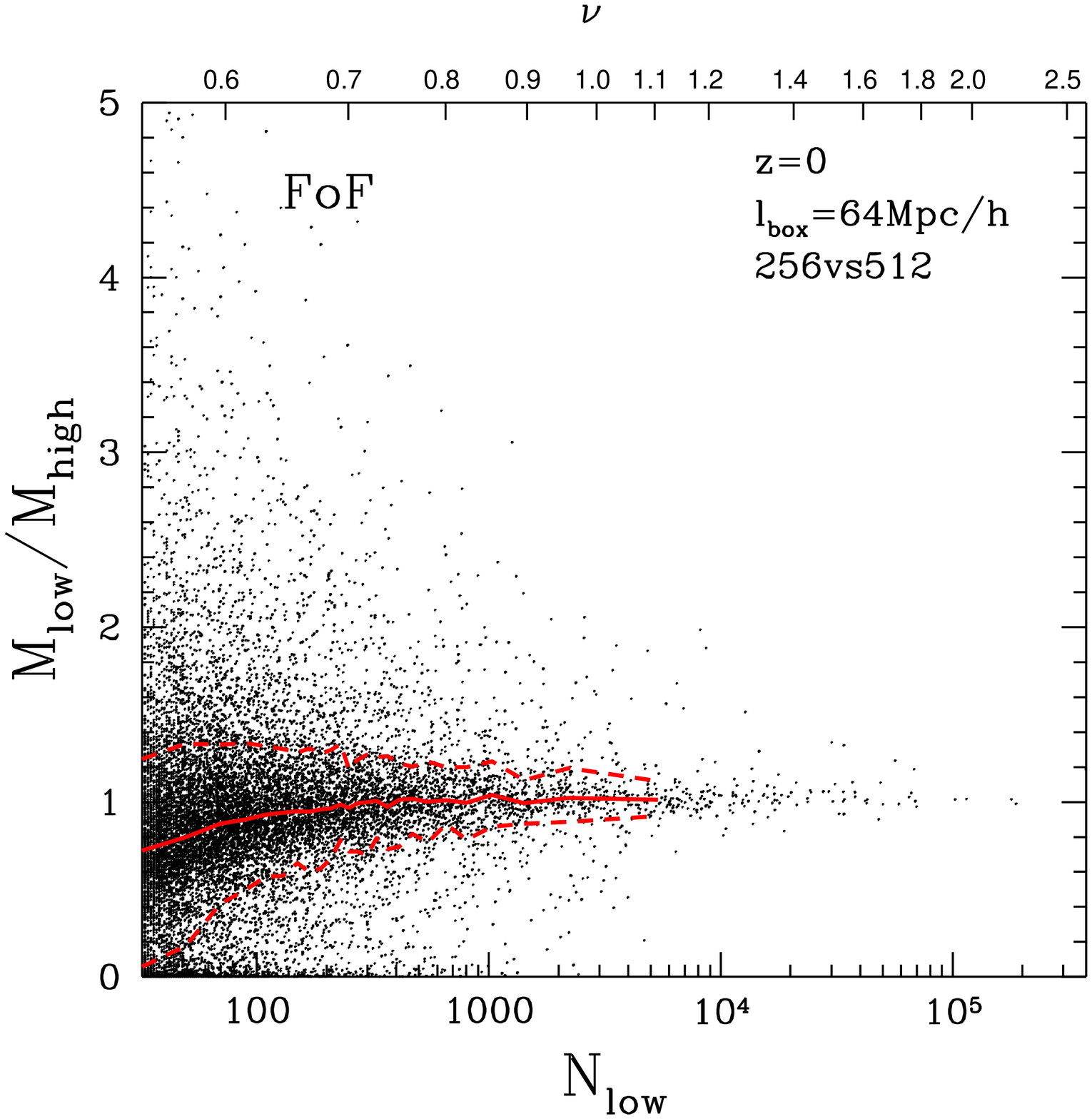}
\includegraphics[scale=0.4]{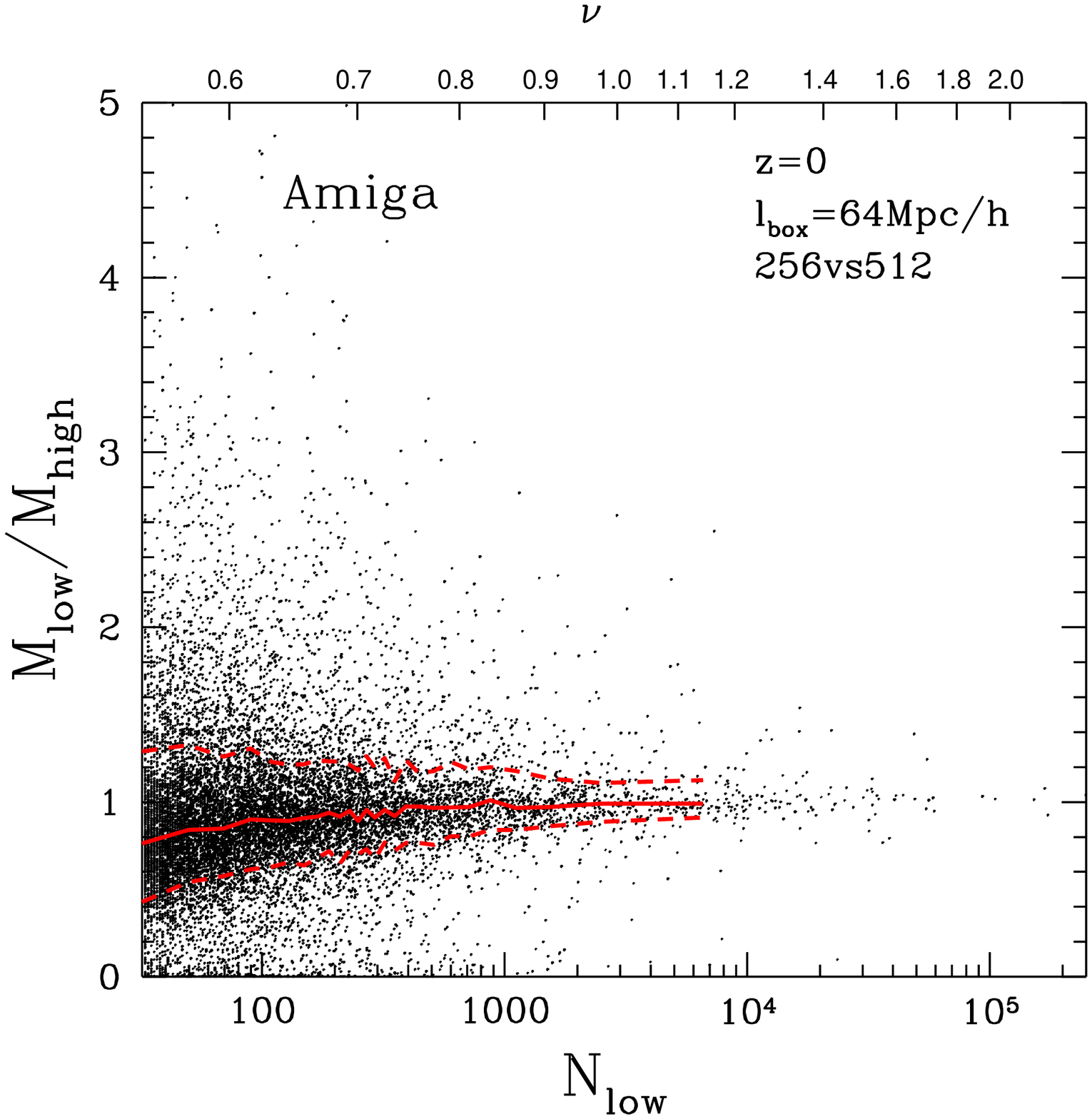}
\caption{Upper left panel: Ratio of low-to-high resolution mass of
  dark matter halos identified in the $N_p=128^3$ version of our
  medium box ($l_{box}=64~\mathrm{Mpc~h^{-1}}$) at $z=0$ and compared
  to higher resolution realization ($N_p=512^3$) with same
  initial conditions.  The scatter plot is shown as function of
  number of low-resolution particles.  Solid red line represents
  median value of ratio, and red dashed lines enclose
  $1\sigma$ symmetric confidence region around median. The upper
  axis translates halo particle number into dimensionless
  variable $\nu=\delta^2_c/\sigma^2(M)$, a measure of the rarity of a
  halo in the extended Press-Schechter formalism. Upper right panel:
  same as upper left panel, but here dark-matter halos have been
  identified with Amiga halo finder \citep{knebe08} including
  removal of unbound particles. Lower left and right panels are
  same as the upper set, but for $N_p=256^3$ run compared to
  $N_p=512^3$ run.}
\label{fig:mass_scatter} \end{figure}

\begin{figure} \includegraphics[scale=0.9]{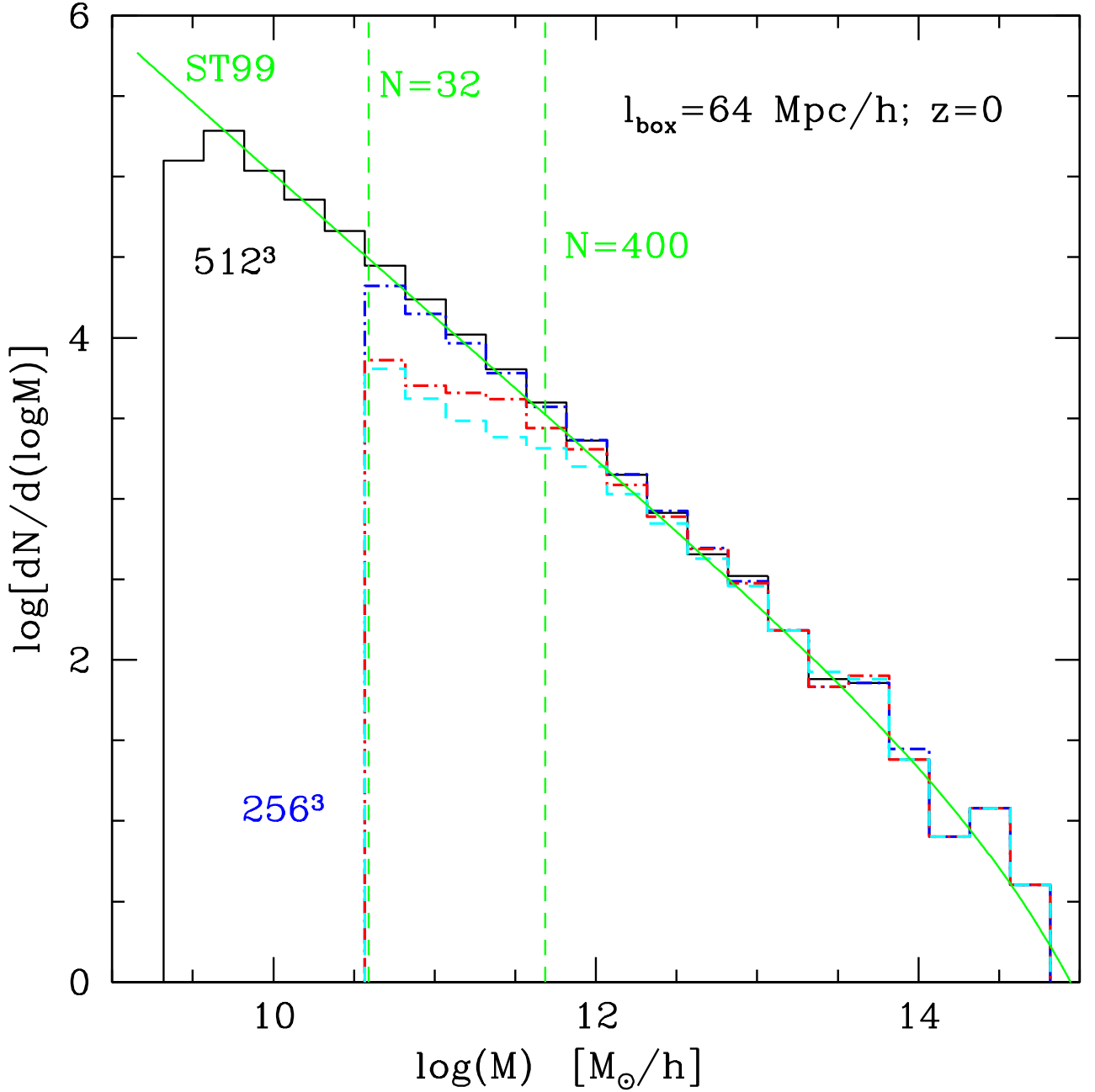} \caption{Halo mass
    function at $z=0$ of medium box simulated with $N_p=512^3$ particles
    (black solid line) compared to $N_p=256^3$ realizations
    (Gadget2  blue dotted-dashed line; Enzo AMR red dotted-dashed line; Enzo
    Unigrid cyan dashed line). For comparison, we also plot
\citet{st99} halo mass function (solid green line).
 The low-resolution halo mass
 function obtained in the Gadget2 run is consistent with
 high-resolution and analytical mass function down to its resolution
 limit ($N \sim 30$). The halo mass function from the Enzo runs
 converges only when $N\gtrsim 400$.}
\label{fig:mf} \end{figure}

\begin{figure} \includegraphics[scale=0.4]{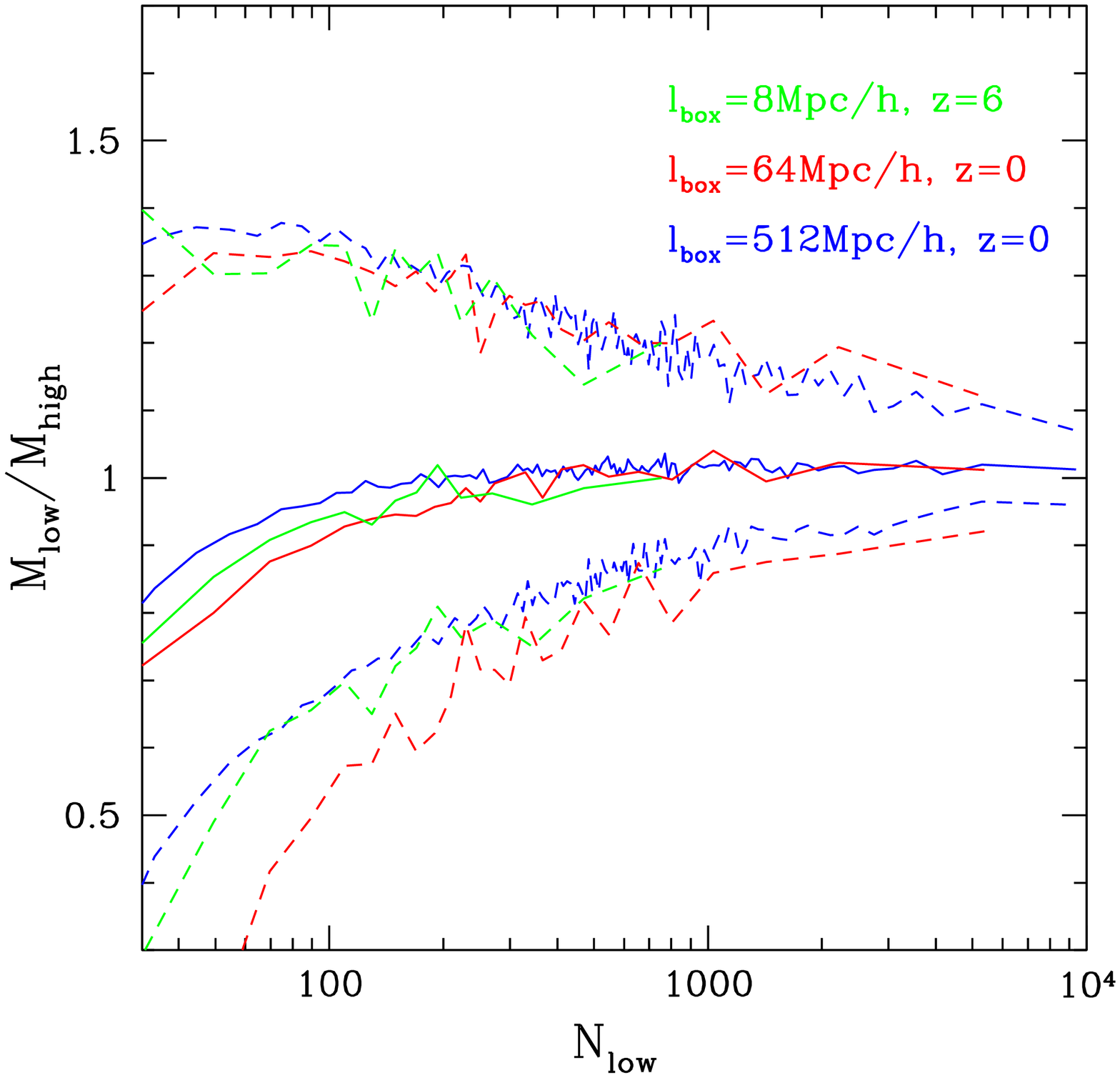}
\includegraphics[scale=0.4]{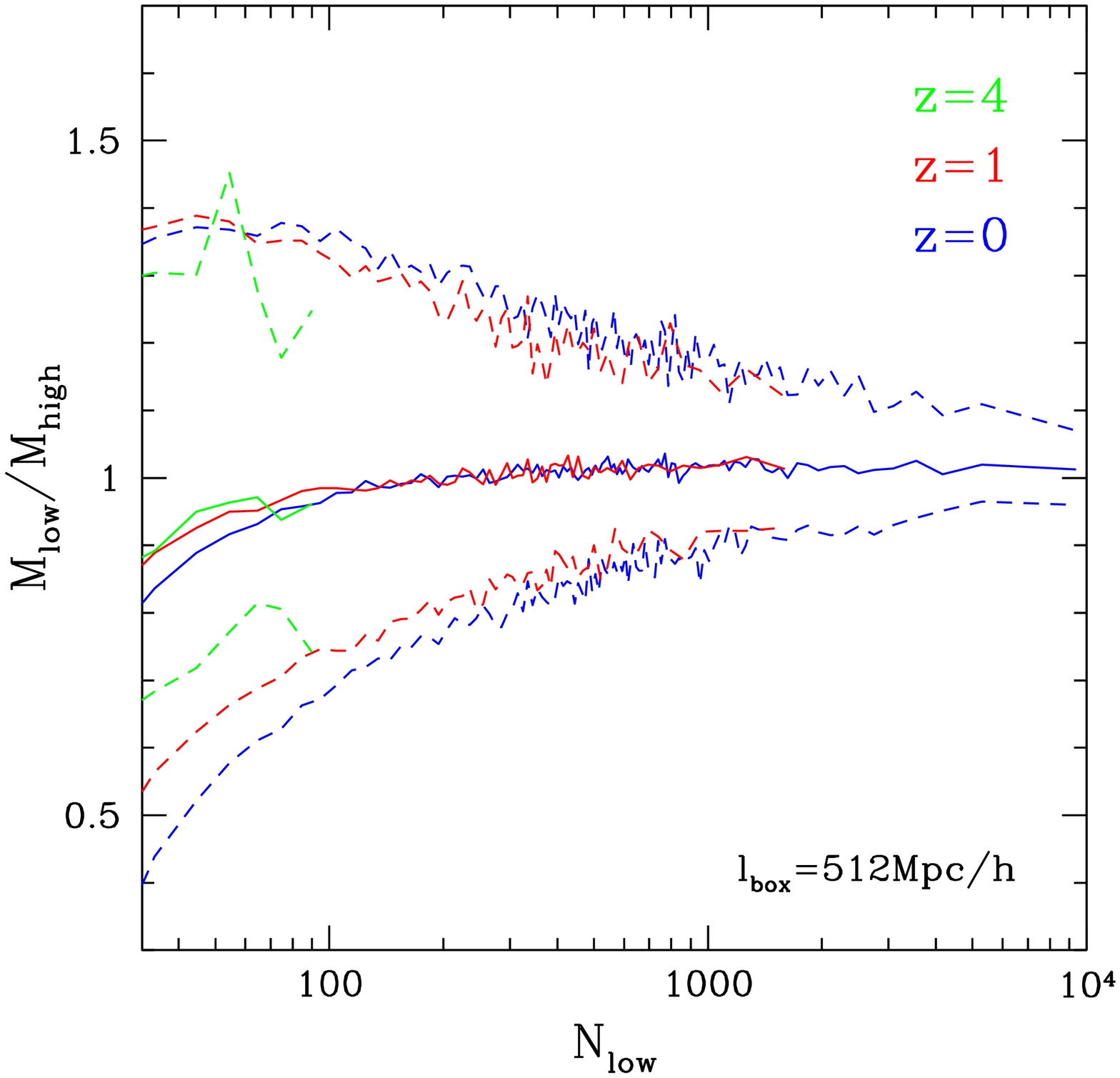} 
\caption{{ $1\sigma$ confidence intervals (dashed lines) and median
  (solid lines) for the ratio of low-to-high resolution mass of dark
  matter halos as in Figure~\ref{fig:mass_scatter} but considering
  different simulations. The left panel investigates different box
  sizes. Blue lines are associated with the $N_{low}=512^3$ version of
  our large box ($l_{box}=512~\mathrm{Mpc~h^{-1}}$) compared to the
  $N_{high}=1024^3$ realization at $z=0$. Red lines refer to
  $N_{low}=256^3$ version of medium box
  ($l_{box}=64~\mathrm{Mpc~h^{-1}}$) compared to $N_{high}=512^3$
  realization at $z=0$. Green lines refer to $N_{low}=256^3$ version
  of small box ($l_{box}=64~\mathrm{Mpc~h^{-1}}$) compared to
  $N_{high}=512^3$ realization at $z=6$. Right panel investigates the
  redshift dependence of convergence for large box realization
  $N_{low}=512^3$ compared to $N_{high}=1024^3$ at $z=0$ (blue), $z=1$
  (red), and $z=4$ (green). Halos at fixed number of particles are
  rarer at higher redshift and have slightly less scatter in the
  mass measurements.}}
 \label{fig:mass_scatter5} \end{figure}

In this paper we quantify the convergence of individual halo
properties, as the resolution of a cosmological simulation is
increased while maintaining the same large-scale structure of the
coarser run. We confirm past investigations of the convergence of the
global mass function down to our resolution limit of $N=32$ particles
in runs carried out with the particle-mesh tree code Gadget2. At the
same time, we demonstrate that the scatter in individual halo masses
measured from our set of cosmological simulations is qualitatively and
quantitatively different from that reported based on the analysis of
the Friends-of Friends halo finder on mock halos and on downsampled
snapshots of an individual cosmological simulation
\citep{warren06,lukic09}. Those experiments highlighted the tendency
of the FoF halo finder to \emph{overestimate} the mass of poorly
resolved halos. We show instead that, in an actual resolution study of
cosmological simulations, the mass of low-N halos tends to be
\emph{underestimated}. This behavior in the convergence of low N halo
properties is determined primarily by the finite accuracy of the
gravity integration, rather than by the properties of the FoF
algorithm explored in \citet{warren06,lukic09}.  In fact, the results
obtained in Figure~\ref{fig:mass_scatter} with the Amiga halo finder
\citep{knebe08} are fully consistent with those obtained with the FoF
finder, despite the fact that Amiga identifies halos from the topology
of isodensity contours.

\begin{figure} 
\includegraphics[scale=0.9]{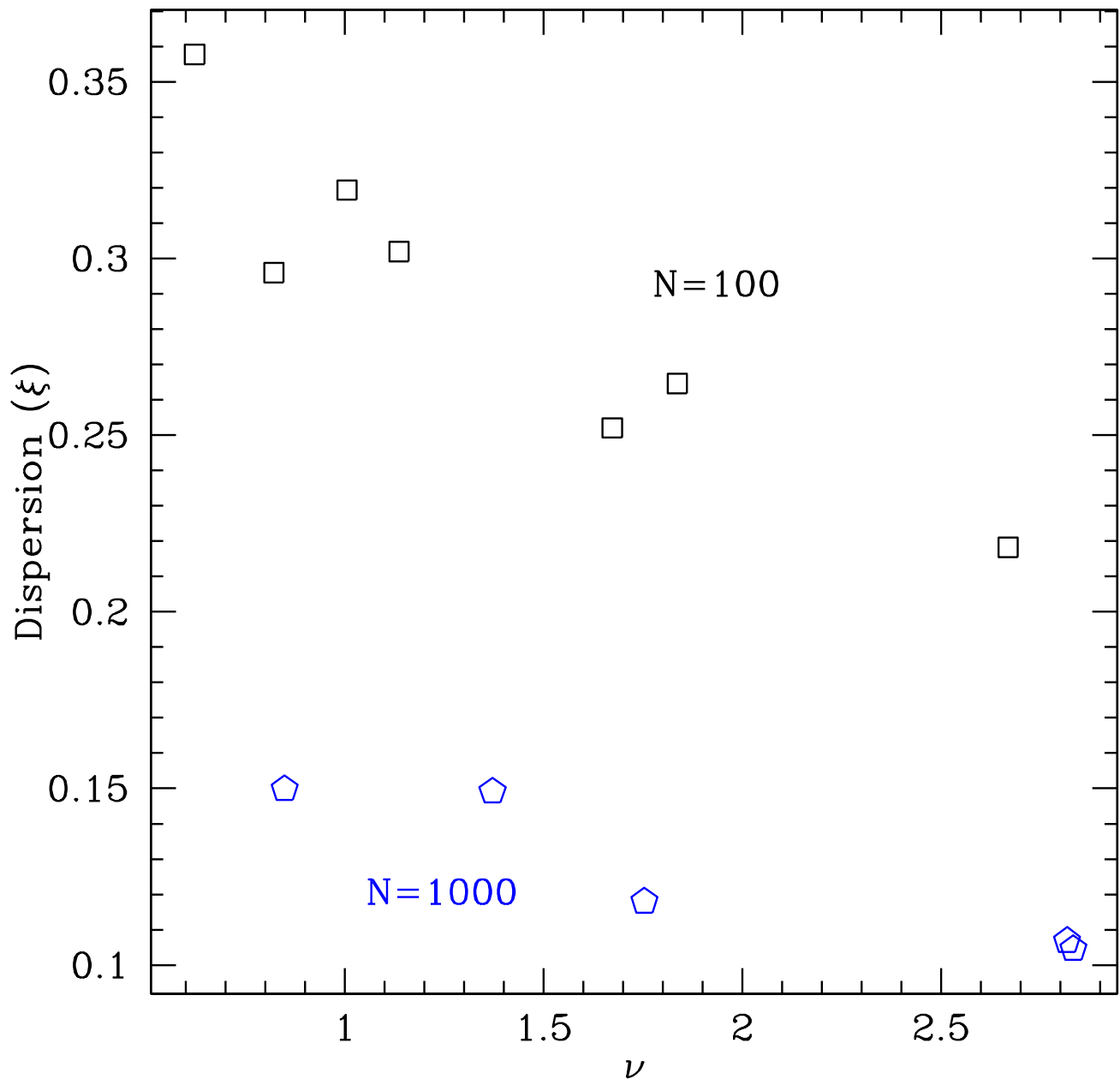}
\caption{The $1\sigma$ dispersion $\xi$ around median for
  $M_{low}/M_{high} $ distribution in our set of simulations
  considering halos with $100 \leq N_{low} \leq 200$ (black squares)
  and with $1000 \leq N_{low} \leq 2000$ (blue pentagons) as function
  of rarity $\nu$. Rare (large $\nu$) halos resolved with a low number
  of particles have less scatter than their more common counterparts.}
\label{fig:sigma_nu} \end{figure}

\begin{figure} 
\includegraphics[scale=0.9]{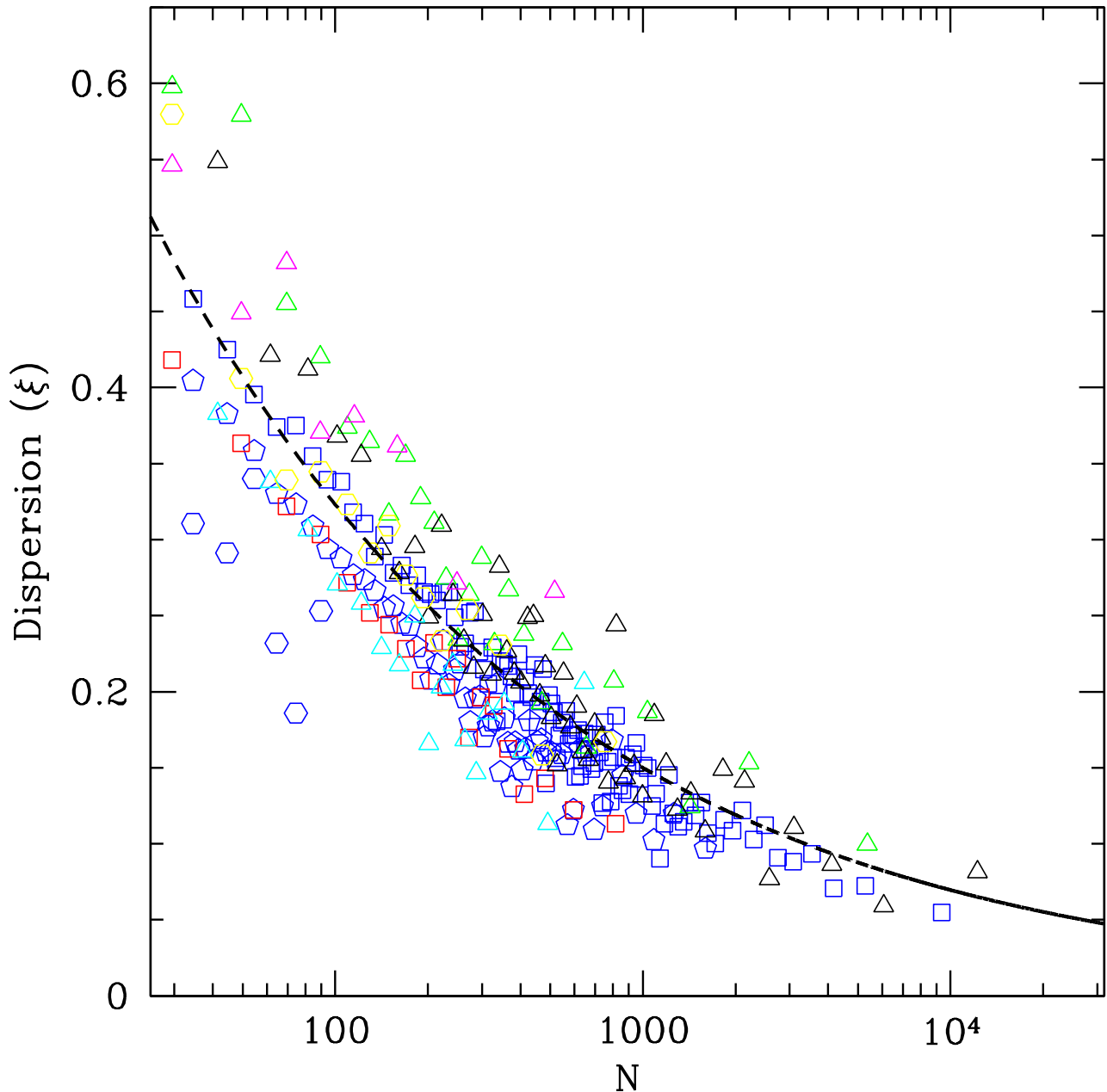}
\caption{The $1\sigma$ dispersion $\xi$ around median for
  $M_{low}/M_{high} $ distribution in our set of simulations as
  function of number of particles in low-resolution run. Triangles
  show color-coded results from medium box: common halos in the
  256vs512 snapshot at $z=0$ (green) and at $z=1$ (black), rarer halos
  in the $z=4$ snapshot (light blue) and halos in the 128vs512
  snapshot at $z=0$ (violet).  Squares are from large box at $z=0$
  (red 256vs1024; blue 512vs1024).  Pentagons are from large box
  (512vs1024) at $z=1$ and blue hexagons are from same simulation at
  $z=4$, representing halos with smallest $\xi(N)$ (due to their
  rarity).  Yellow hexagons are from small box at $z=6$ (256vs512).
  Dashed line represents the analytical scaling predicted in
  Section~\ref{sec:result} ($\xi \propto N^{-1/3}$).}
\label{fig:sigma_npart} \end{figure}

\begin{figure} 
\includegraphics[scale=0.52]{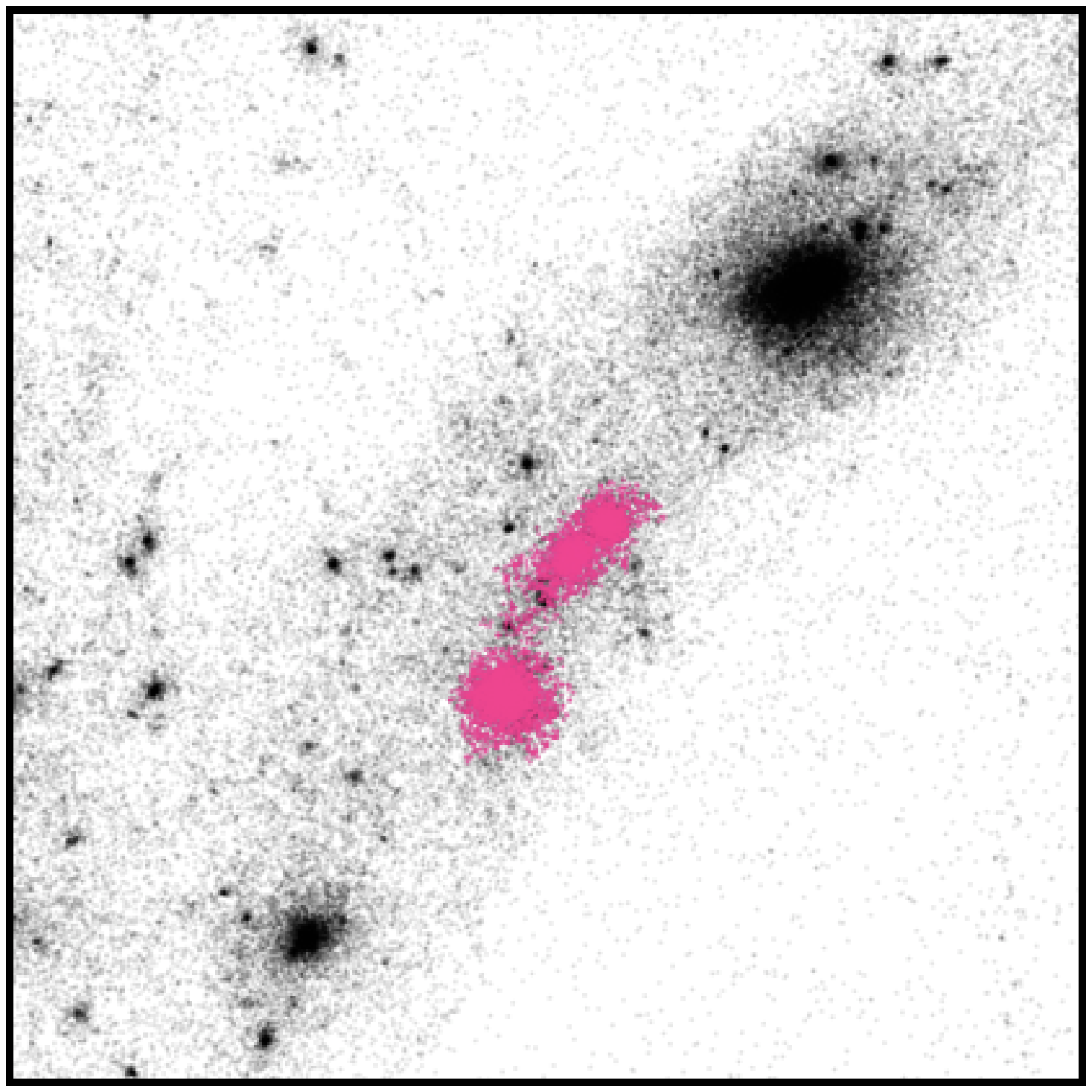}
\includegraphics[scale=0.52]{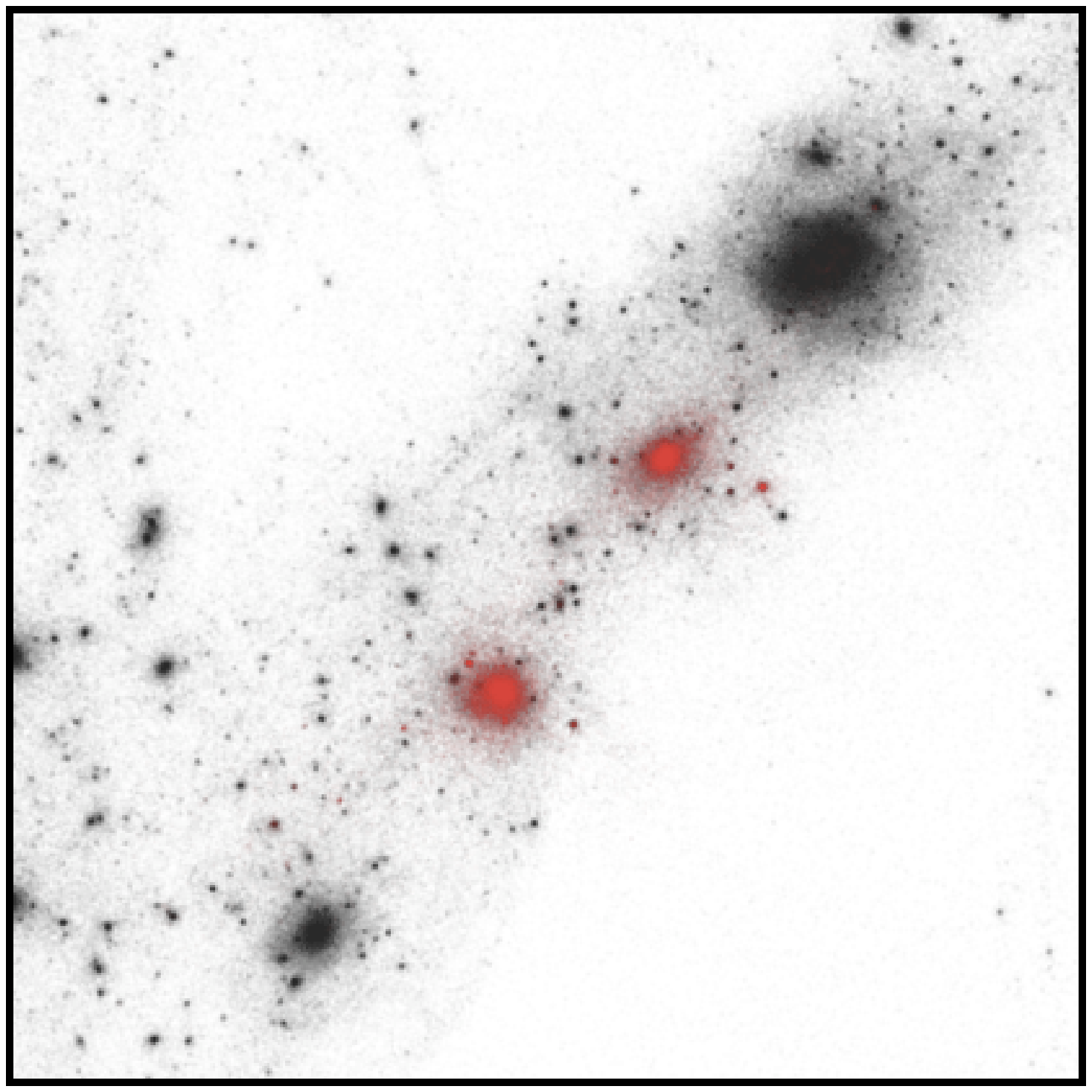}
\caption{$(7 ~\mathrm{Mpc}~{h^{-1}})^3$ region around one halo of our
  medium-box simulations at low resolution (left $N_p=256^3$) and high
  resolution (right $N_p = 512^3$). The 8192 particles belonging to
  the halo at low resolution are shown in magenta (left), while their
  counterparts at high resolution are in red (right). This halo has
  two distinct counterparts at high resolution, because the merging is
  slightly delayed compared to low-resolution realization, where a
  thin bridge of particles already connects the two sub-components.}
  \label{fig:halo_merger} \end{figure}

In addition to the different direction of the systematic bias, the
scatter in the individual halo masses is more than a factor two higher
than reported previously for idealized halos. The mass of halos
resolved with  $N\sim 1000$ particles shows a $\sim 15\%$ scatter when
the halos are resimulated at higher ($8\times$) resolution. Halos with
a smaller number of particles have a larger uncertainty in their
masses, typically of the order of $\sim 30$\% for $N\sim100$. The
$1\sigma$ relative uncertainty in the mass of a single halo with $N$
particles scales approximately as $\xi \sim N^{-1/3}$ (see
Eq.~\ref{eq:xi} and Figure~\ref{fig:sigma_npart}). As the resolution
limit of the simulation is approached, halo masses show less scatter
if the small-$N$ halos are rare, and thus the simulation box is mostly
composed of particles still near the regime of linear evolution. The
mass of rarer halos can instead be measured relatively well, even with
a low number of particles (see Fig.~\ref{fig:sigma_nu}). The core
density and the spin of small-$N$ halos exhibit a systematic bias in
addition to significant scatter. The virial radius of a halo appears to be
the quantity with the smallest scatter, no bias down to
$N=32$.

\begin{figure} \includegraphics[scale=0.4]{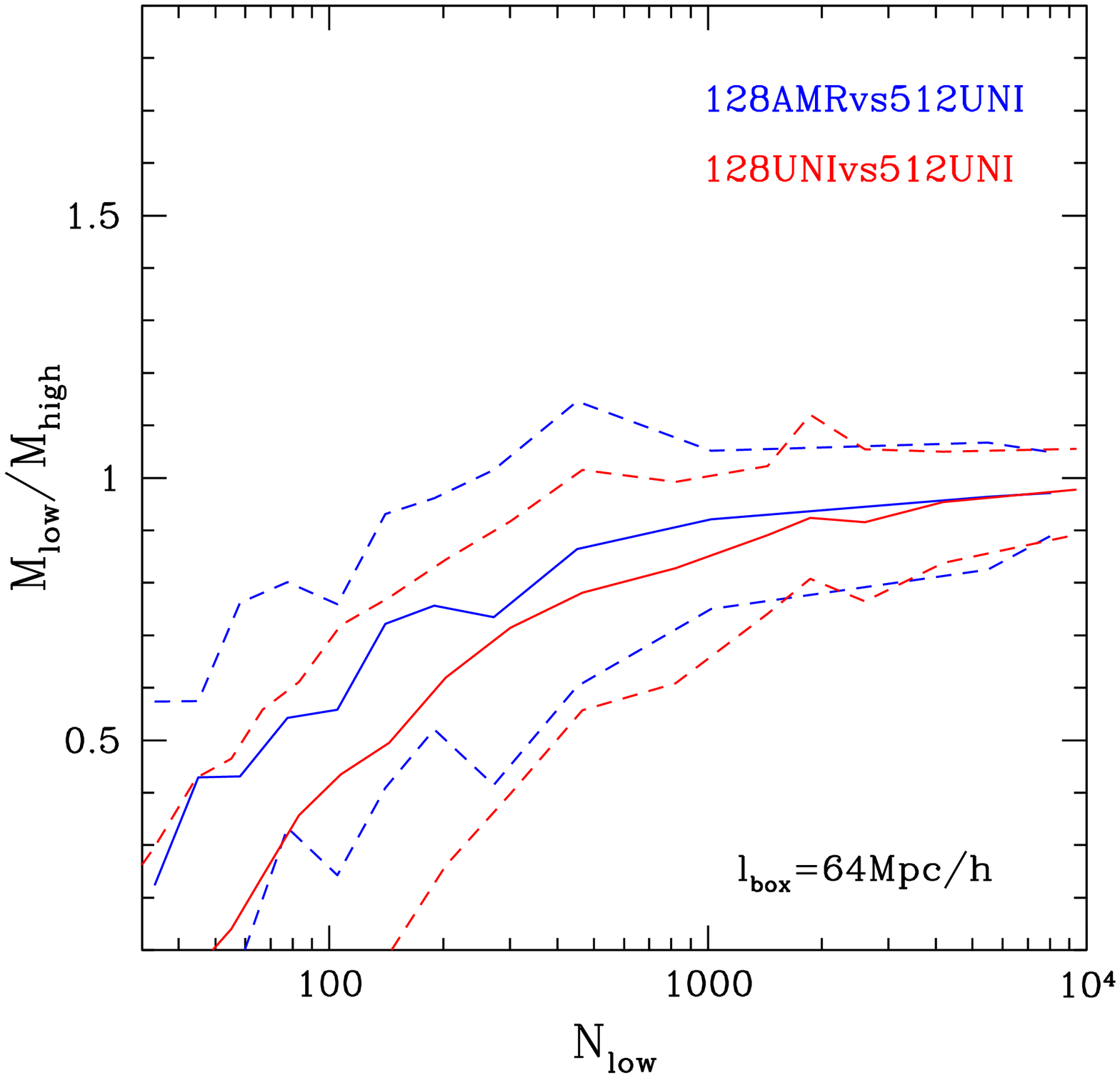}
 \includegraphics[scale=0.4]{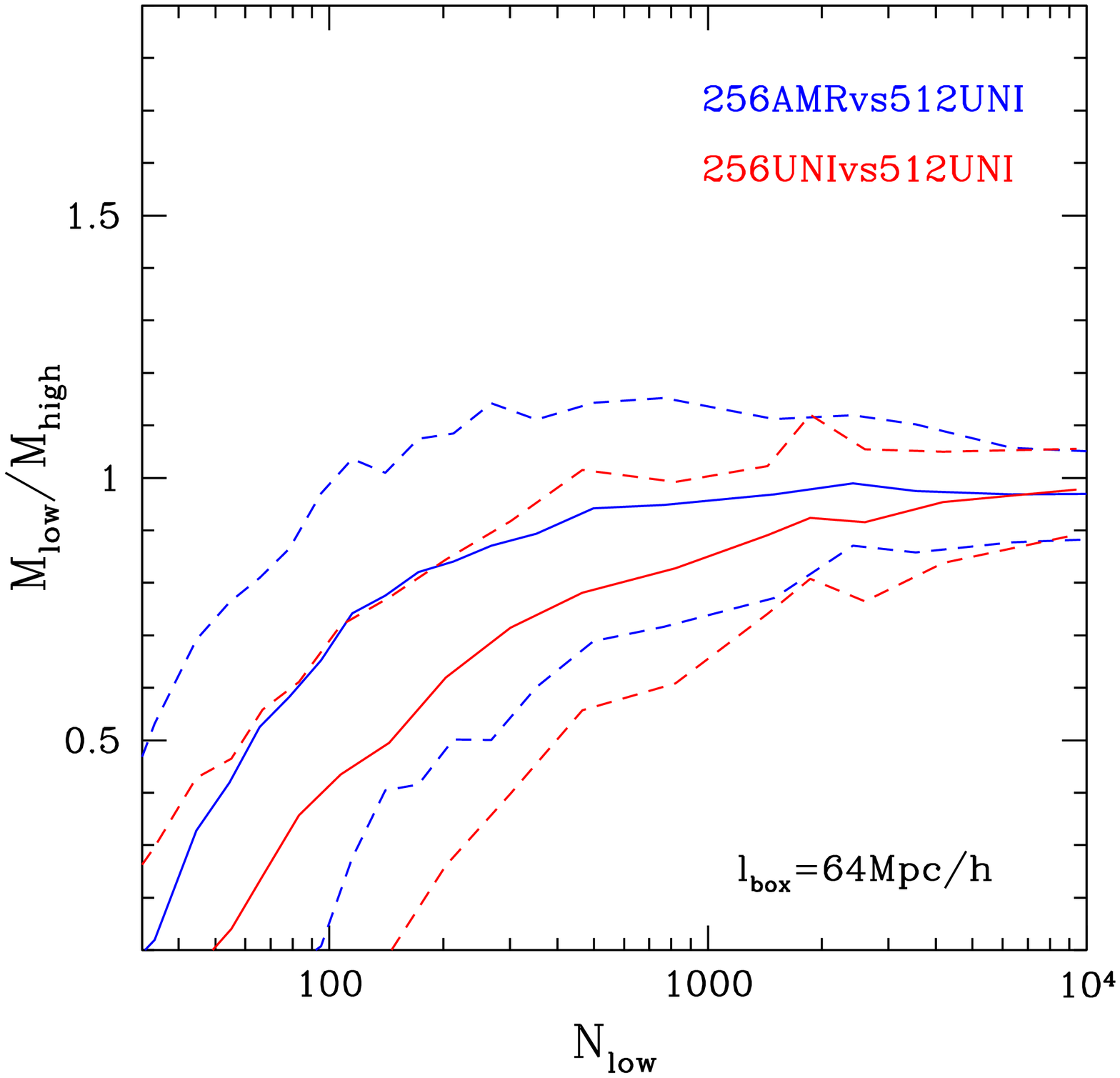}
\caption{{ Same as Fig.~\ref{fig:mass_scatter5} but for
   runs carried out with the Enzo code. The left panel compares $N_{low}=128^3$ in 
Unigrid mode, shown as red lines, and in AMR mode (6 levels), shown as blue lines, to $N_{high}=512^3$ (Unigrid) at $z=0$. 
The right panel is as the left one but for $N_{low}=256^3$.
The mass of individual halos with $N \lesssim 500$ is
  significantly underestimated in Enzo. AMR helps to improve
  convergence, but its effectiveness is limited.}}
 \label{fig:mass_scatter_enzo} \end{figure}

We carried out our analysis using a code --- Gadget2 --- that can
efficiently reach a high force resolution (a small fraction of the
initial inter-particle distance) at low computational cost.
Simulations done with the AMR code Enzo are prone to a systematic bias
in individual halo mass when $N \lesssim 1000$. Resorting to Adaptive
Mesh Refinement lessens but does not resolve this problem (see
Figure~\ref{fig:mass_scatter_enzo}). Increasing the top-grid dimension
helps to improve convergence of individual halo masses with $N\sim
100$ (see \citealt{oshea05}).

Our investigation highlights the importance of carrying out careful
resolution studies to validate the conclusions of numerical
simulations. This is particularly important when sub-grid physics
recipes are implemented within a numerical simulation or when
analytical models of galaxy formation are constructed from the
simulation snapshots. If these recipes prescribe to populate halos
with a small number of particles, one might obtain the correct average
behavior of the sample (because the halo-mass function converges down
to halos with $\sim 30$ particles). However, the limited numerical
resolution is likely to introduce extra scatter in the properties of
the end-products of such simulations. This extra scatter may also
propagate from low-$N$ progenitors to descendant halos (and galaxies) of
larger mass, for instance if the star-formation recipes depend
critically on the initial metal enrichment. 

\begin{figure}
\includegraphics[scale=0.4]{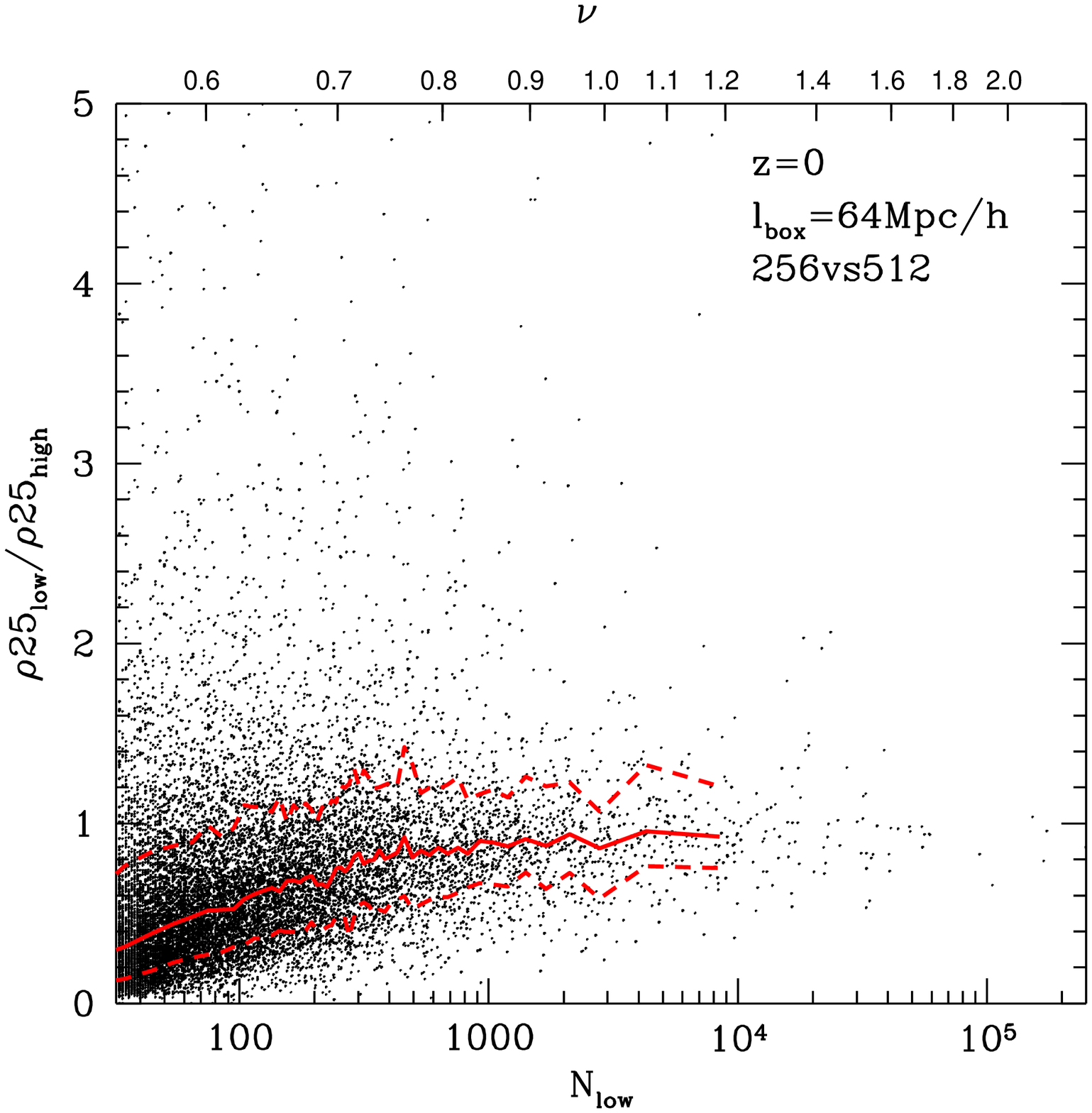}
\includegraphics[scale=0.4]{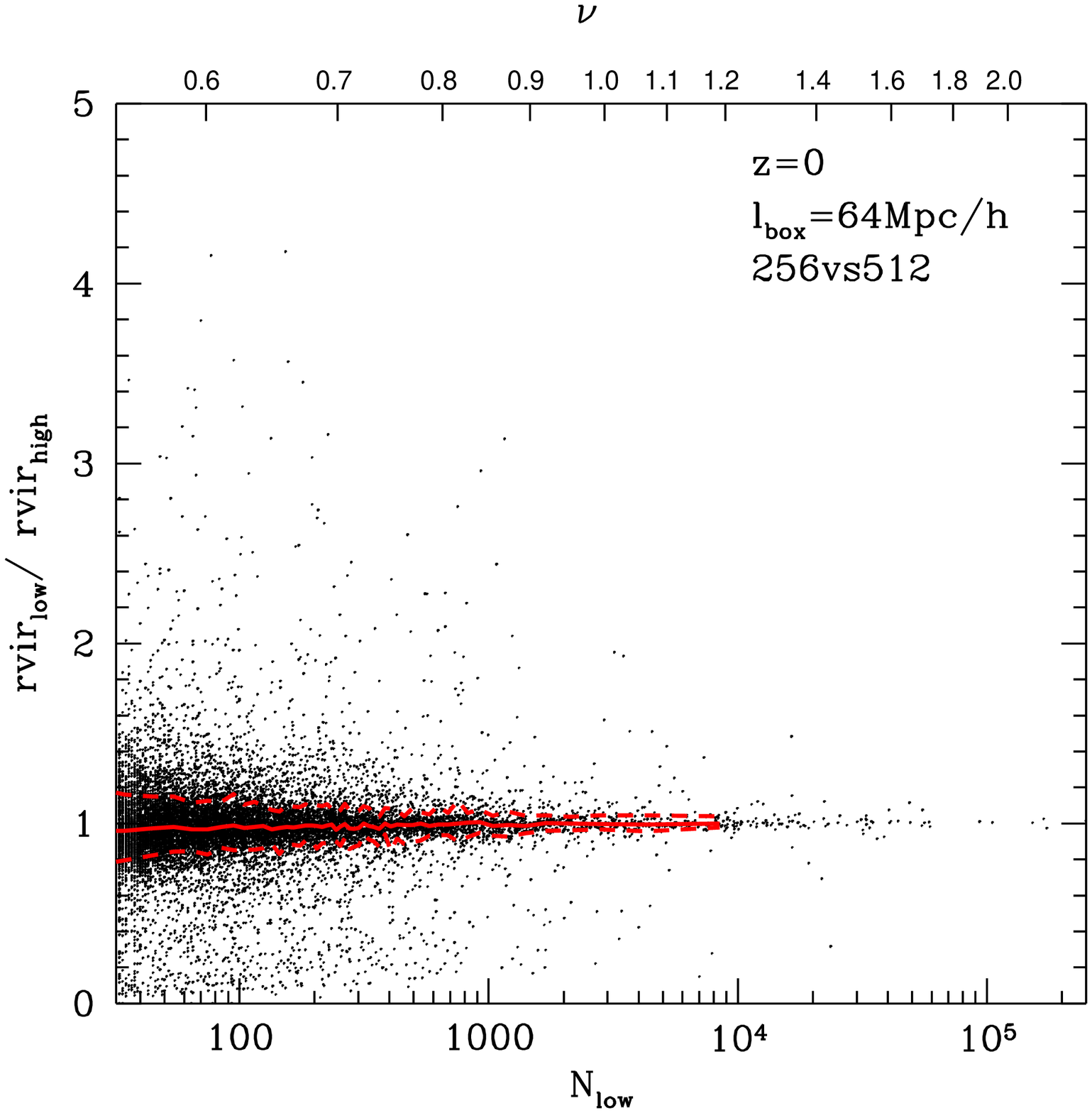}
\includegraphics[scale=0.4]{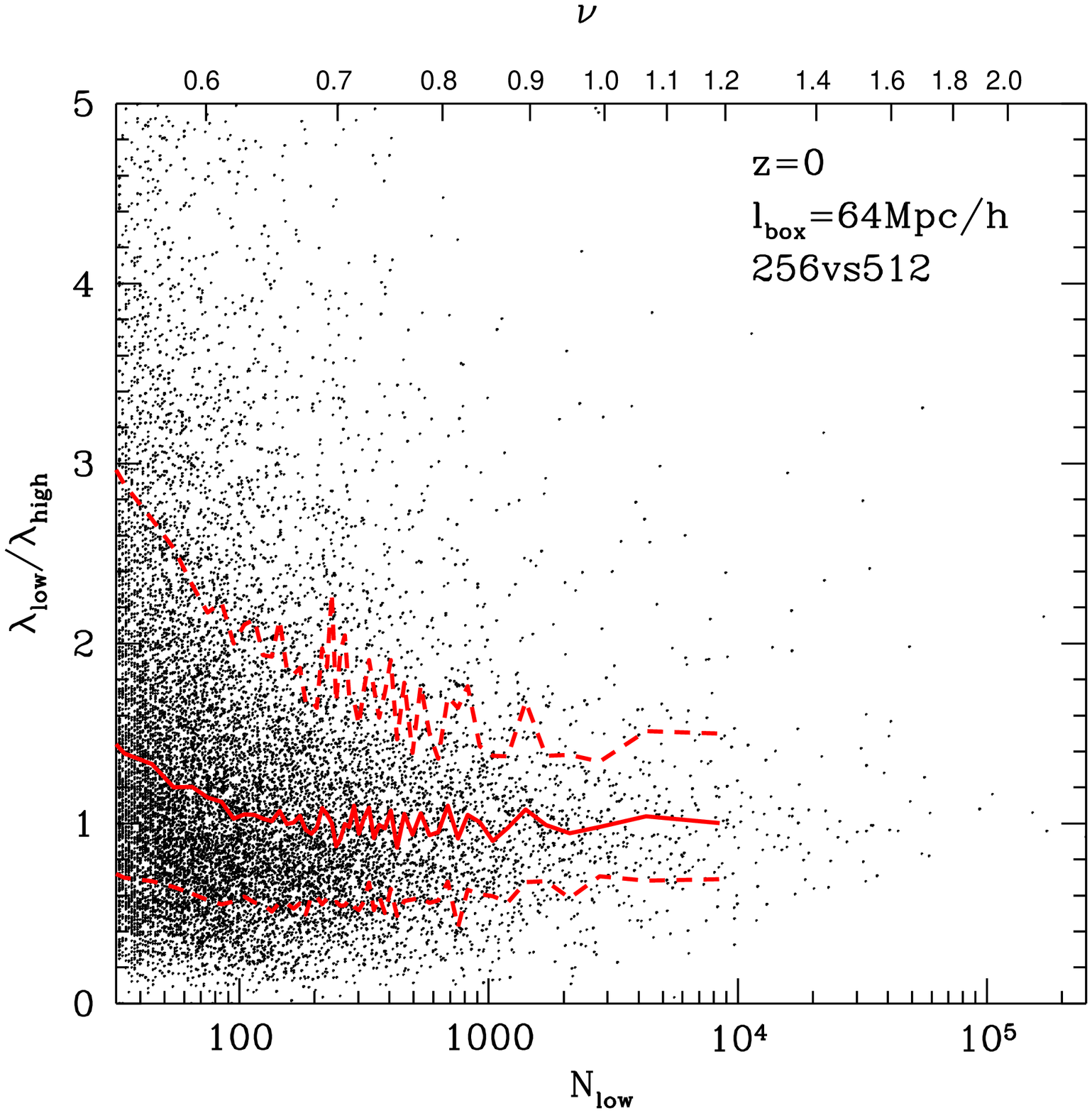}
\caption{{Convergence of additional properties of dark matter halos  in $N_p=256^3$ realization
    of medium box compared to their high-resolution counterparts
    ($N_p=512^3$ run).  Upper-left panel: Ratio of core density $\rho_{25~low}/\rho_{25~high}$. Upper-right panel: Ratio of virial radius $r_{vir~low}/r_{vir~high}$. Lower left panel: Ratio of dimensionless angular momentum
    $\lambda_{low}/\lambda_{high}$. For each panel the solid red line shows the median of distribution. Dashed
    red lines delimit the symmetric $1\sigma$ region around median.}}
\label{fig:rho25}\label{fig:rvir}\label{fig:spin}  \end{figure}

Furthermore, systematic biases are possible. One obvious example is
the rate of star formation calculated from the central density of a
halo.  Figure~\ref{fig:rho25} clearly shows that the core density of
halos with $N \lesssim 400$ is underestimated by more than $20\%$ (and
by more than $75\%$ at $N\sim 32$). Even recipes based on halo mass
alone can lead to systematic biases. One example is the ratio of
Ly-$\alpha$ to stellar luminosity in a simple model where the stellar
luminosity is proportional to the halo mass, while the Ly-$\alpha$
luminosity depends on $M_{HI}^2 \propto M^2$ \citep{dayal09}. The
non-linear relation of Ly-$\alpha$ luminosity on the halo mass is
therefore affected by the scatter in $M_{low}/M_{high}$. For example,
in our medium-box 256vs512 simulations at $z=0$, $\langle
M_{low}/M_{high} \rangle \sim 0.84$ while $\langle
(M_{low}/M_{high})^2 \rangle \sim 1.33$ for halos with $32 \leq N \leq
64$. This means that the Ly-$\alpha$ luminosity would be overestimated
compared to the stellar luminosity by a factor $\sim 1.6$ at low
resolution.
  
Another example is given by the convergence study carried out in
\citet{trenti09b} to validate their simulations for the transition
from metal-free to metal-enriched star formation during the
reionization epoch. They found that a ``low'' resolution run
($N_p=512^3$) achieved full convergence with the high-resolution run
($N_p=1024^3$) only at $z \lesssim 7$, when sub-grid physics was
implemented in DM halos with $N\gtrsim 100$ particles (see
Fig.~6 in \citealt{trenti09b}). Individual scatter in halo masses
might also introduce numerical noise when feedback is considered. For
example, if supernova feedback from star formation at the center of
the halos is near the critical level to evacuate most of the baryons,
then the numerical uncertainty in the halo mass might play a critical
role for the future development of the star formation history in the
descendant halos of poorly resolved progenitors.

Overall we recommend implementing extra physics only in halos with $N
\gtrsim 100$ for rare halos ($\nu \gtrsim 3$) and $N \gtrsim 400$ for
very common halos ($\nu \lesssim 1$). This will guarantee $\xi(N)
\lesssim 0.2$ and adequate convergence of other halo properties. If
this recommendation is followed, then both the Enzo code and Gadget2
are in a regime where their convergence properties are similar. Of
course, many interesting problems in cosmology require a greater
dynamic range than is currently possible to resolve. Thus,
implementing extra physics only on halos resolved with $N \gtrsim 100$
particles is not optimal. In such cases, modelers should take
precautions to demonstrate convergence of their halo simulations with
$N \lesssim 100-400$ and avoid extrapolating beyond the range of
validity.

\acknowledgements 

{ We thank Zarija Luki{\'c}, Brian O'Shea and Darren Reed for
  stimulating comments, Steffen Knollmann for his advice on compiling
  and running the Amiga halo finder code, and
    an anonymous referee for useful suggestions}. We acknowledge
  support from the University of Colorado Astrophysical Theory Program
  through grants from NASA (NNX07AG77G) and NSF (AST07-07474). This
  research was supported in part by the National Science Foundation
  through TeraGrid awards TG-AST090040 and TG-AST090101.

\clearpage

\bibliography{popIII}{}

\begin{thebibliography}{48}
\expandafter\ifx\csname natexlab\endcsname\relax\def\natexlab#1{#1}\fi

\bibitem[{{Bagla} \& {Ray}(2005)}]{bagla05}
{Bagla}, J.~S. \& {Ray}, S. 2005, \mnras, 358, 1076

\bibitem[{{Berger} \& {Colella}(1989)}]{berger1989}
{Berger}, M.~J. \& {Colella}, P. 1989, Journal of Computational Physics, 82, 64

\bibitem[{{Bertschinger}(1998)}]{bert98}
{Bertschinger}, E. 1998, \araa, 36, 599

\bibitem[{{Bertschinger}(2001)}]{bertschinger01}
---. 2001, \apjs, 137, 1

\bibitem[{{Bryan} {et~al.}(1995){Bryan}, {Norman}, {Stone}, {Cen}, \&
  {Ostriker}}]{bryan95}
{Bryan}, G.~L., {Norman}, M.~L., {Stone}, J.~M., {Cen}, R., \& {Ostriker},
  J.~P. 1995, Computer Physics Communications, 89, 149

\bibitem[{{Colella} \& {Woodward}(1984)}]{woodward1984}
{Colella}, P. \& {Woodward}, P.~R. 1984, Journal of Computational Physics, 54,
  174

\bibitem[{{Davis} {et~al.}(1985){Davis}, {Efstathiou}, {Frenk}, \&
  {White}}]{davis85}
{Davis}, M., {Efstathiou}, G., {Frenk}, C.~S., \& {White}, S.~D.~M. 1985, \apj,
  292, 371

\bibitem[{{Dayal} {et~al.}(2009){Dayal}, {Ferrara}, \& {Saro}}]{dayal09}
{Dayal}, P., {Ferrara}, A., \& {Saro}, A. 2009, ArXiv 0907.4989

\bibitem[{{De Lucia} \& {Blaizot}(2007)}]{delucia07}
{De Lucia}, G. \& {Blaizot}, J. 2007, \mnras, 375, 2

\bibitem[{{Diemand} {et~al.}(2007){Diemand}, {Kuhlen}, \& {Madau}}]{vialactea}
{Diemand}, J., {Kuhlen}, M., \& {Madau}, P. 2007, \apj, 657, 262

\bibitem[{{Djorgovski} \& {Davis}(1987)}]{fp87}
{Djorgovski}, S. \& {Davis}, M. 1987, \apj, 313, 59

\bibitem[{{Eisenstein} \& {Hu}(1999)}]{eisenstein99}
{Eisenstein}, D.~J. \& {Hu}, W. 1999, \apj, 511, 5

\bibitem[{{Fan} {et~al.}(2006){Fan}, {Strauss}, \& {Richards}}]{fan06}
{Fan}, X., {Strauss}, M.~A., \& {Richards}, e.~a. 2006, \aj, 131, 1203

\bibitem[{{Ferrarese} \& {Merritt}(2000)}]{ferrarese00}
{Ferrarese}, L. \& {Merritt}, D. 2000, \apjl, 539, L9

\bibitem[{{Fukushige} {et~al.}(2004){Fukushige}, {Kawai}, \&
  {Makino}}]{fukushige04}
{Fukushige}, T., {Kawai}, A., \& {Makino}, J. 2004, \apj, 606, 625

\bibitem[{{Ghigna} {et~al.}(2000){Ghigna}, {Moore}, {Governato}, {Lake},
  {Quinn}, \& {Stadel}}]{ghigna00}
{Ghigna}, S., {Moore}, B., {Governato}, F., {Lake}, G., {Quinn}, T., \&
  {Stadel}, J. 2000, \apj, 544, 616

\bibitem[{{Hallman} {et~al.}(2007){Hallman}, {O'Shea}, {Burns}, {Norman},
  {Harkness}, \& {Wagner}}]{hallman07}
{Hallman}, E.~J., {O'Shea}, B.~W., {Burns}, J.~O., {Norman}, M.~L., {Harkness},
  R., \& {Wagner}, R. 2007, \apj, 671, 27

\bibitem[{{Heitmann} {et~al.}(2006){Heitmann}, {Luki{\'c}}, {Habib}, \&
  {Ricker}}]{heitmann06}
{Heitmann}, K., {Luki{\'c}}, Z., {Habib}, S., \& {Ricker}, P.~M. 2006, \apjl,
  642, L85

\bibitem[{{Hoffman} \& {Ribak}(1991)}]{hoffman91}
{Hoffman}, Y. \& {Ribak}, E. 1991, \apjl, 380, L5

\bibitem[{{Jenkins} {et~al.}(2001){Jenkins}, {Frenk}, {White}, {Colberg},
  {Cole}, {Evrard}, {Couchman}, \& {Yoshida}}]{jenkins01}
{Jenkins}, A., {Frenk}, C.~S., {White}, S.~D.~M., {Colberg}, J.~M., {Cole}, S.,
  {Evrard}, A.~E., {Couchman}, H.~M.~P., \& {Yoshida}, N. 2001, \mnras, 321,
  372

\bibitem[{{Kauffmann} \& {Charlot}(1998)}]{kauffmann98}
{Kauffmann}, G. \& {Charlot}, S. 1998, \mnras, 294, 705

\bibitem[{{Knollmann} \& {Knebe}(2009)}]{knebe08}
{Knollmann}, S.~R. \& {Knebe}, A. 2009, \apjs, 182, 608

\bibitem[{{Komatsu} {et~al.}(2009){Komatsu}, {Dunkley}, \& {Nolta}}]{komatsu08}
{Komatsu}, E., {Dunkley}, J., \& {Nolta}, M.~R. e.~a. 2009, \apjs, 180, 330

\bibitem[{{Luki{\'c}} {et~al.}(2007){Luki{\'c}}, {Heitmann}, {Habib},
  {Bashinsky}, \& {Ricker}}]{lukic07}
{Luki{\'c}}, Z., {Heitmann}, K., {Habib}, S., {Bashinsky}, S., \& {Ricker},
  P.~M. 2007, \apj, 671, 1160

\bibitem[{{Luki{\'c}} {et~al.}(2009){Luki{\'c}}, {Reed}, {Habib}, \&
  {Heitmann}}]{lukic09}
{Luki{\'c}}, Z., {Reed}, D., {Habib}, S., \& {Heitmann}, K. 2009, \apj, 692,
  217

\bibitem[{{Norman} {et~al.}(2009){Norman}, {Paschos}, \& {Harkness}}]{unigrid0}
{Norman}, M.~L., {Paschos}, P., \& {Harkness}, R. 2009, Journal of Physics
  Conference Series, 180, 012021

\bibitem[{{Navarro} {et~al.}(1997){Navarro}, {Frenk}, \& {White}}]{NFW}
{Navarro}, J.~F., {Frenk}, C.~S., \& {White}, S.~D.~M. 1997, \apj, 490, 493

\bibitem[{{Navarro} {et~al.}(2010){Navarro}, {Ludlow}, {Springel}, {Wang},
  {Vogelsberger}, {White}, {Jenkins}, {Frenk}, \& {Helmi}}]{navarro08}
{Navarro}, J.~F., {Ludlow}, A., {Springel}, V., {Wang}, J., {Vogelsberger}, M.,
  {White}, S.~D.~M., {Jenkins}, A., {Frenk}, C.~S., \& {Helmi}, A. 2010, \mnras, in press

\bibitem[{{Oppenheimer} \& {Dav{\'e}}(2008)}]{oppen08}
{Oppenheimer}, B.~D. \& {Dav{\'e}}, R. 2008, \mnras, 387, 577

\bibitem[{{O'Shea} {et~al.}(2005){O'Shea}, {Nagamine}, {Springel}, {Hernquist},
  \& {Norman}}]{oshea05}
{O'Shea}, B.~W., {Nagamine}, K., {Springel}, V., {Hernquist}, L., \& {Norman},
  M.~L. 2005, \apjs, 160, 1

\bibitem[{{O'Shea} \& {Norman}(2007)}]{oshea07}
{O'Shea}, B.~W. \& {Norman}, M.~L. 2007, \apj, 654, 66


\bibitem[{{Paschos} {et~al.}(2009){Paschos}, {Jena}, {Tytler}, {Kirkman}, \&
  {Norman}}]{unigrid2}
{Paschos}, P., {Jena}, T., {Tytler}, D., {Kirkman}, D., \& {Norman}, M.~L.
  2009, \mnras, 399, 1934


\bibitem[{{Power} {et~al.}(2003){Power}, {Navarro}, {Jenkins}, {Frenk},
  {White}, {Springel}, {Stadel}, \& {Quinn}}]{power03}
{Power}, C., {Navarro}, J.~F., {Jenkins}, A., {Frenk}, C.~S., {White},
  S.~D.~M., {Springel}, V., {Stadel}, J., \& {Quinn}, T. 2003, \mnras, 338, 14

\bibitem[{{Reed} {et~al.}(2003){Reed}, {Gardner}, {Quinn}, {Stadel}, {Fardal},
  {Lake}, \& {Governato}}]{reed03}
{Reed}, D., {Gardner}, J., {Quinn}, T., {Stadel}, J., {Fardal}, M., {Lake}, G.,
  \& {Governato}, F. 2003, \mnras, 346, 565

\bibitem[{{Reed} {et~al.}(2007){Reed}, {Bower}, {Frenk}, {Jenkins}, \&
  {Theuns}}]{reed07}
{Reed}, D.~S., {Bower}, R., {Frenk}, C.~S., {Jenkins}, A., \& {Theuns}, T.
  2007, \mnras, 374, 2

\bibitem[{{Reed} {et~al.}(2009){Reed}, {Bower}, {Frenk}, {Jenkins}, \&
  {Theuns}}]{reed09}
---. 2009, \mnras, 394, 624

\bibitem[{{Regan} {et~al.}(2007){Regan}, {Haehnelt}, \& {Viel}}]{unigrid3}
{Regan}, J.~A., {Haehnelt}, M.~G., \& {Viel}, M. 2007, \mnras, 374, 196

\bibitem[{{Schmidt}(1959)}]{schmidt59}
{Schmidt}, M. 1959, \apj, 129, 243

\bibitem[{{Sheth} {et~al.}(2001){Sheth}, {Mo}, \& {Tormen}}]{sheth01}
{Sheth}, R.~K., {Mo}, H.~J., \& {Tormen}, G. 2001, \mnras, 323, 1

\bibitem[{{Sheth} \& {Tormen}(1999)}]{st99}
{Sheth}, R.~K. \& {Tormen}, G. 1999, \mnras, 308, 119

\bibitem[{{Somerville} \& {Primack}(1999)}]{somerville99}
{Somerville}, R.~S. \& {Primack}, J.~R. 1999, \mnras, 310, 1087

\bibitem[{{Springel}(2005)}]{springel05}
{Springel}, V. 2005, \mnras, 364, 1105

\bibitem[{{Springel} \& {Hernquist}(2003)}]{springel03}
{Springel}, V. \& {Hernquist}, L. 2003, \mnras, 339, 289

\bibitem[{{Springel} {et~al.}(2008a){Springel}, {Wang}, {Vogelsberger},
  {Ludlow}, {Jenkins}, {Helmi}, {Navarro}, {Frenk}, \& {White}}]{aquarius}
{Springel}, V., {Wang}, J., {Vogelsberger}, M., {Ludlow}, A., {Jenkins}, A.,
  {Helmi}, A., {Navarro}, J.~F., {Frenk}, C.~S., \& {White}, S.~D.~M. 2008,
  \mnras, 391, 1685

\bibitem[Springel et al.(2008b)]{springel08b} Springel et al. 2008, \nat, 456, 73

\bibitem[{{Springel} {et~al.}(2005){Springel}, {White}, \& {Jenkins}}]{mill}
{Springel}, V., {White}, S.~D.~M., \& {Jenkins}, A. e.~a. 2005, \nat, 435, 629

\bibitem[{{Stadel} {et~al.}(2009){Stadel}, {Potter}, {Moore}, {Diemand},
  {Madau}, {Zemp}, {Kuhlen}, \& {Quilis}}]{stadel09}
{Stadel}, J., {Potter}, D., {Moore}, B., {Diemand}, J., {Madau}, P., {Zemp},
  M., {Kuhlen}, M., \& {Quilis}, V. 2009, \mnras, 398, L21

\bibitem[{{Trenti} \& {Stiavelli}(2007)}]{ts07a}
{Trenti}, M. \& {Stiavelli}, M. 2007, \apj, 667, 38

\bibitem[{{Trenti} {et~al.}(2008){Trenti}, {Santos}, \& {Stiavelli}}]{tss08}
{Trenti}, M., {Santos}, M.~R., \& {Stiavelli}, M. 2008, \apj, 687, 1

\bibitem[{{Trenti} {et~al.}(2009){Trenti}, {Stiavelli}, \& {Shull}}]{trenti09b}
{Trenti}, M., {Stiavelli}, M., \& {Shull}, J.~M. 2009, \apj, 700, 1672

\bibitem[{{Turk} {et~al.}(2009){Turk}, {Abel}, \& {O'Shea}}]{turk09}
{Turk}, M.~J., {Abel}, T., \& {O'Shea}, B. 2009, Science, 325, 601

\bibitem[{{Tytler} {et~al.}(2009){Tytler}, {Paschos}, {Kirkman}, {Norman}, \&
  {Jena}}]{unigrid1}
{Tytler}, D., {Paschos}, P., {Kirkman}, D., {Norman}, M.~L., \& {Jena}, T.
  2009, \mnras, 393, 723

\bibitem[{{Warren} {et~al.}(2006){Warren}, {Abazajian}, {Holz}, \&
  {Teodoro}}]{warren06}
{Warren}, M.~S., {Abazajian}, K., {Holz}, D.~E., \& {Teodoro}, L. 2006, \apj,
  646, 881

\bibitem[{{Yepes} {et~al.}(2008){Yepes}, {Martinez-Vaquero}, {Khalatyan},
  {Gottl{\"o}ber}, {Hoeft}, \& {Teyssier}}]{marenostrum}
{Yepes}, G., {Martinez-Vaquero}, L., {Khalatyan}, A., {Gottl{\"o}ber}, S.,
  {Hoeft}, M., \& {Teyssier}, R. 2008, in IAU Symposium, Vol. 244, IAU
  Symposium, ed. {J.~Davies \& M.~Disney}, 393--394

\end{thebibliography}

\clearpage

\begin{table} \begin{center} \caption{Simulation Summary\label{tab:sim}}
\smallskip
\smallskip
\begin{tabular}{ccccc}
\tableline\tableline
$N_p^{1/3}$ & $l_{box}$ & $m_{DM}$ &$z_{start}$ & $z_{end}$ \\
\tableline

\smallskip  128 & 512 Mpc h$^{-1}$ & $4.98 \times 10^{12} ~{M_{\sun}~h^{-1}}$&100 & 0\\
\smallskip 256 & 512 Mpc h$^{-1}$ & $6.22 \times 10^{11} ~{M_{\sun}~h^{-1}}$& 100 & 0\\
\smallskip 512 & 512 Mpc h$^{-1}$ & $7.78 \times 10^{10} ~{M_{\sun}~h^{-1}}$& 100 & 0\\
\smallskip 1024 & 512 Mpc h$^{-1}$ & $9.73 \times 10^{9}~ {M_{\sun}~h^{-1}}$& 100 &0\\
\tableline
\smallskip  64 & 64 Mpc h$^{-1}$ & $7.78 \times 10^{10} ~{M_{\sun}~h^{-1}}$& 100& 0 \\
\smallskip 128 & 64 Mpc h$^{-1}$ & $9.73 \times 10^{9} ~{M_{\sun}~h^{-1}}$ & 100&0 \\
\smallskip 256 & 64 Mpc h$^{-1}$ &$1.21 \times 10^{9} ~{M_{\sun}~h^{-1}}$ & 100&0 \\
\smallskip 512 & 64 Mpc h$^{-1}$ &$1.52 \times 10^{8} ~{M_{\sun}~h^{-1}}$ & 100&0 \\
\tableline
\smallskip 128 & 8 Mpc h$^{-1}$ & $1.90 \times 10^{7} ~{M_{\sun}~h^{-1}}$& 199&6 \\
\smallskip 256 & 8 Mpc h$^{-1}$ &$2.37 \times 10^{6} ~{M_{\sun}~h^{-1}}$ & 199&6 \\
\smallskip 512 & 8 Mpc h$^{-1}$ &$2.97 \times 10^{5} ~{M_{\sun}~h^{-1}}$ & 199&6 \\
\tableline
\end{tabular}
\tablecomments{Summary of the properties of our cosmological
  simulations done with Gadget2. The first column reports the number
  of dark matter particles $N_p$, the second the box-size
  $l_{box}$. The single-particle dark-matter mass ($m_{DM}$) is in the third
  column, the initial redshift $z_{start}$ in the fourth,
  and the final redshift $z_{end}$ in the last column. We also carried
  out the medium-box runs ($l_{box}= 64~\mathrm{Mpc}~{h^{-1}}$) with the
  Enzo code, in both Unigrid and AMR (6 levels) mode for runs up to
  $N_p=256^3$ and in Unigrid mode only for $N_p=512^3$. }
\end{center}
\end{table}

\clearpage

\end{document}